\documentclass{aa}
\usepackage{graphics}
\usepackage{txfonts}
\usepackage{xcolor}
\usepackage{changepage}
\usepackage{blindtext}
\begin{document}
\title{Formation and Dynamics of Transequatorial Loops}
\author{Avyarthana Ghosh \and Durgesh Tripathi}
\institute{Inter-University Centre for Astronomy and Astrophysics, Post Bag - 4, Ganeshkhind, Pune 411007, India\\
\email{avyarthana@iucaa.in ; durgesh@iucaa.in}}
\abstract{}{To study the dynamical evolution of trans-equatorial loops (TELs) using imaging and spectroscopy.}{We have used the images recorded by the Atmospheric Imaging Assembly and the Helioseismic Magnetic Imager on-board the Solar Dynamics Observatory and spectroscopic observations taken from the Extreme-Ultraviolet Imaging Spectrometer on-board Hinode.}{The data from AIA 193~{\AA} channel show that TELs are formed between AR 12230 and a newly emerging AR 12234 and evolved during December 10{--}14, 2014. The xt-plots for December 12, 2014 obtained using AIA 193~{\AA} data reveal signatures of inflow and outflow towards an X-region. High cadence AIA images also show recurrent intensity enhancements in close proximity to the X-region (P2), which is observed to have higher intensities for spectral lines formed at $\log\,T[K] = $6.20 and voids at other higher temperatures. The electron densities and temperatures in the X-region (and P2) are maintained steadily at $\log\,N_e = $8.5{--}8.7~cm$^{-3}$ and $\log\,T[K] = $6.20, respectively. Doppler velocities in the X-region show predominant redshifts by about 5{--}8~km~s$^{-1}$ when closer to the disk centre but blueshifts (along with some zero-velocity pixels) when away from the centre. The Full-Width-Half-Maxima (FWHM) maps reveal non-thermal velocities of about 27{--}30~km~s$^{-1}$ for \ion{Fe}{xii}, \ion{Fe}{xiii} and \ion{Fe}{xv} lines. However, the brightest pixels have non- thermal velocities $\sim$62~km~s$^{-1}$ for \ion{Fe}{xii} and \ion{Fe}{xiii} lines. On the contrary, the dark X-region for \ion{Fe}{xv} line have the highest non-thermal velocity ($\sim$115~km~s$^{-1}$).}{We conclude that the TELs are formed due to magnetic reconnection. We further note that the TELs themselves undergo magnetic reconnection leading to reformation of loops of individual ARs. Moreover, this study, for the first time, provides measurements of plasma parameters in X-regions thereby providing essential constraints for theoretical studies. }


\maketitle


\section{Introduction}
It is well established that the solar corona is full of loop structures \citep[see, e.g.,][for a review]{Rea_2014} as well as diffuse emissions \citep{DelM_2003, TriMD_2009, ViaK_2011, SubTK_2014}. One of those are the trans-equatorial loops (TELs), which join ARs across the equator of the Sun. The theoretical existence of such loop structures was first suggested by \cite{Bab_1961} as a consequence of the solar dynamo. However, such loops were first reported over a decade later in 1974 using the observations recorded by the X-Ray Telescope (XRT) on-board Skylab \citep{VaiKP_1974}. Based on the subsequent observations of ARs associated with McMath plage numbers 12472 and 12474, \cite{ChaKS_1976} and \cite{SveKC_1977} suggested that the TELs were formed due to magnetic reconnection between two ARs. 

Using the observations recorded by the Soft X-ray Telescope \citep[][]{TsuAB_1991} on-board \textsl{Yohkoh}, \cite{Tsu_1996} reported the formation of X-type topology in TELs connecting two ARs in two hemispheres. This has been followed by a number of authors \citep{Far_1999, Pev_2000, FarKS_2001, FarS_2002, CroGK_2002, Pev_2004, CheBZ_2006, CheBZ_2007, WanZZ_2007, ShiNK_2007, YokM_2009, YokM_2010} reporting the formation, characteristics and evolution of TELs, using the observations taken from SXT as well as the X-Ray Telescope \citep[XRT;][]{GolDA_2007} on-board Hinode \citep{KosMS_2007}. 

The early spectroscopic diagnostics of TELs were reported by authors $viz.,$ \cite{HarMD_2003} and \cite{Bro_2006} using the observations recorded by the Coronal Diagnostic Spectrometer \citep[CDS;][]{HarSC_1995} on-board the Solar and Heliospheric Observatory \citep[SOHO;][]{DomFP_1995}. However, we emphasize that these spectroscopic measurements were performed in and along the off-limb TELs only.  

X-shaped regions (cusp regions) are believed to be signatures of magnetic reconnection, for $e.g.,$ \cite{TsuHS_1992, ForA_1996,TriSS_2006, TriSMW_2007}. Therefore, measurement of plasma parameters in the X-shaped structure formed between the TELs may help us probe the physical properties of reconnection regions. To the best of our knowledge, there are no such measurements till date for TELs.

In this paper, we perform a detailed study of a complete sequence of the formation and evolution of a set of TELs using the observations taken from the Atmospheric Imaging Assembly \citep[AIA;][]{Lem_2012} and the Helioseismic Magnetic Imager \citep[HMI;][]{SchBN_2012, SchSB_2012} on-board the Solar Dynamics Observatory \citep[SDO;][]{BoeEL_2012, PesTC_2012}. This is accompanied with spectroscopic observations recorded with the Extreme-ultraviolet Imaging Spectrometer \citep[EIS;][]{CulHJ_2007} on-board Hinode. The rest of the paper is structured as follows. In \S\ref{obs}, we discuss the observations used for this study. Data analysis and results are presented in \S\ref{res}. Finally, we summarize and discuss the results in \S\ref{summary}.

\section{Observations and data}\label{obs} 

\begin{figure*}[!htbp]
\centering
\includegraphics[trim=1.cm 0.cm 0.cm 0.cm,width=0.85\textwidth]{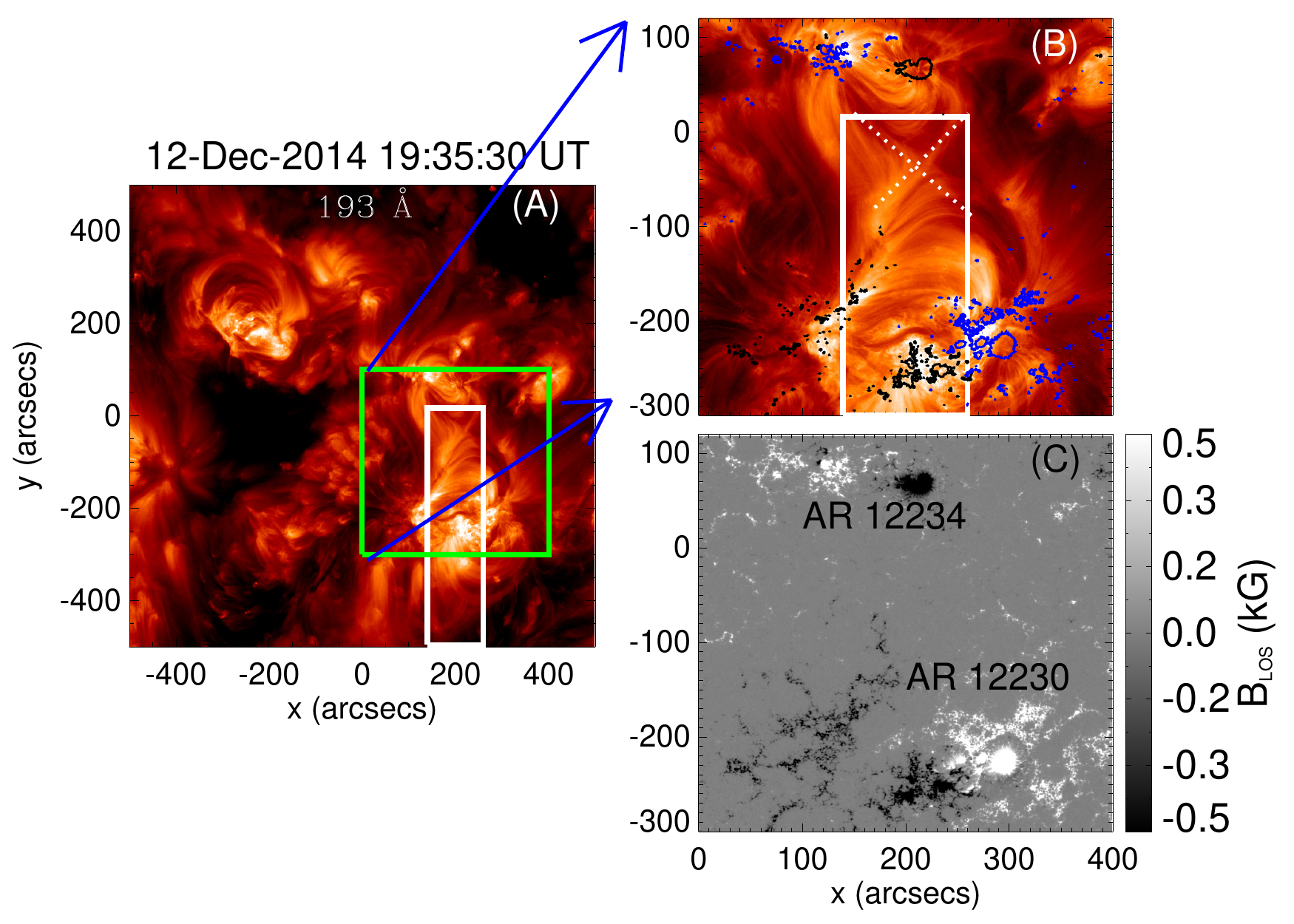}
\caption{Panel (A): AIA 193~{\AA} images showing the portion of the solar disk that contains the two active regions between which the trans-equatorial loops are formed. The over-plotted green box shows the FOV of AIA that is considered for further analysis. The white box is the FOV that was rastered using EIS. Panel (B): Zoom-in view of AIA 193~{\AA} images corresponding to the green box in the left panel showing the X-type topology formed between the TELs (white dotted line). The white box shows the EIS raster FOV. The white dotted lines are drawn to guide the eye on the X-shaped structure. The blue and black contours represent positive and negative polarity regions, respectively with magnetic flux density of $\pm$500~G as observed by HMI at 19:35:40~UT. Panel (C): HMI LOS magnetogram image at the same instant as shown in the contours of panel B. The two active regions \textsl{AR~12230} and \textsl{AR~12234} are marked here.}\label{context}
\end{figure*}

In this work, we have primarily used the AIA and EIS observations. AIA provides near-simultaneous full-disk observations of the solar atmosphere in 7 EUV channels sensitive to different temperatures \citep{DwyDM_2010, DelOM_2011, BoeEL_2012} with an approximate time cadence of 12~s and a pixel size of 0.6{\arcsec}. EIS provides spectroscopic observations in two wavelength bands, CCD~A (170{--}210~{\AA}) and CCD~B (250{--}290~{\AA}) using four slit widths~1, 2, 4 and 250{\arcsec}. We have used the AIA observations to study the long term evolution of the TELs and EIS observations to measure the physical plasma parameters such as electron density, temperature, Doppler velocities and non-thermal widths.

A portion of the solar disk observed on the AIA 193~{\AA} channel on December 12, 2014 at 19:35:30~UT, is shown in panel A of Fig.~\ref{context}. The over-plotted green box indicates the region-of-interest (ROI) and displayed in panel B and considered for further study. The over-plotted white box in both panels A \& B shows the EIS raster field-of-view (FOV). In panel B, we overplot two white dashed lines crossing each other to highlight the X-shaped structure. Fortuitously, the top part of the EIS raster FOV covers this X-shaped region. Also, the blue (positive) and black (negative) contours represent the Line-Of-Sight (LOS) magnetic flux density of 500~G as observed at a near-simultaneous time (19:35:40~UT) by HMI. The same HMI image is shown in panel (C) where we clearly see two bipolar active regions (\textsl{AR~12230} and \textsl{AR~12234}) on either side of the solar equator.

We note that EIS rastered this region four times during 10:16:00{--}11:17:00~UT, 19:04:33{--}20:05:25~UT, 20:40:00{--}21:41:30~UT and 22:16:00{--}23:17:00~UT on December 12th, 2014 using the 2{\arcsec} slit. We denote these rasters as `E1', `E2', `E3' and `E4', respectively, where E2 is the closest to the disk center. All these rasters are 60-steps rasters that covered a FOV of 120\arcsec$\times$512$\arcsec$. We have processed all the AIA and EIS observations with the standard software provided in \textsl{Solarsoft} \citep[][]{FreH_1998}.

\begin{table}[h!]
\caption{EIS spectral lines used for studying the plasma parameters at the reconnection region, as well as the adjacent loops where $\lambda_{0}$ is the rest wavelength. The peak formation temperatures are taken from CHIANTI \citep{DerML_1996, DelDY_2015} at one particular density.}\label{eis_lines}
\begin{center}
\begin{tabular}{| c | c | c |} \hline
Ion name   							&$\lambda_{0}~({\AA})$  			&$\log\,$T$_{peak}$ \\ 
            									& \citep{BroFS_2008} 				&[K]\\ 
\hline					
\ion{Fe}{viii}					&185.213$^{t, w}$ 								&5.80\\
\ion{Si}{vii} 					&275.368$^{w, **}$  						&5.80\\
\ion{Fe}{ix} 				&197.862$^{t}$ 									&5.90\\
\ion{Fe}{x}				&184.536$^{t}$ 									&6.05\\
\ion{Fe}{xi}			      &180.401$^{a,t}$ , 182.167$^{b}$ 					&6.10\\
\ion{Fe}{xii} 			&186.880$^{a}$, 192.394$^{t}$ ,195.119$^{b,t}$ 		&6.20\\
\ion{Fe}{xiii} 			&202.044$^{a,t}$, 203.826$^{b}$						&6.25\\
\ion{Fe}{xiv} 			&264.787$^{a,t}$,270.519$^{b}$  						&6.30\\
\ion{Fe}{xv} 							&284.160$^{t}$							&6.35\\
\hline
\end{tabular}
\end{center} 
$^{a,b}$\textit{density sensitive line pair} \\
$^{t}$\textit{used for temperature diagnostics} \\
$^{w}$\textit{used for deriving the reference wavelength}\\
$^{**}$\textit{Taken from \cite{WarUY_2011}}
\end{table}

\begin{figure*}[!htbp]
\centering
\includegraphics[width=1.0\textwidth]{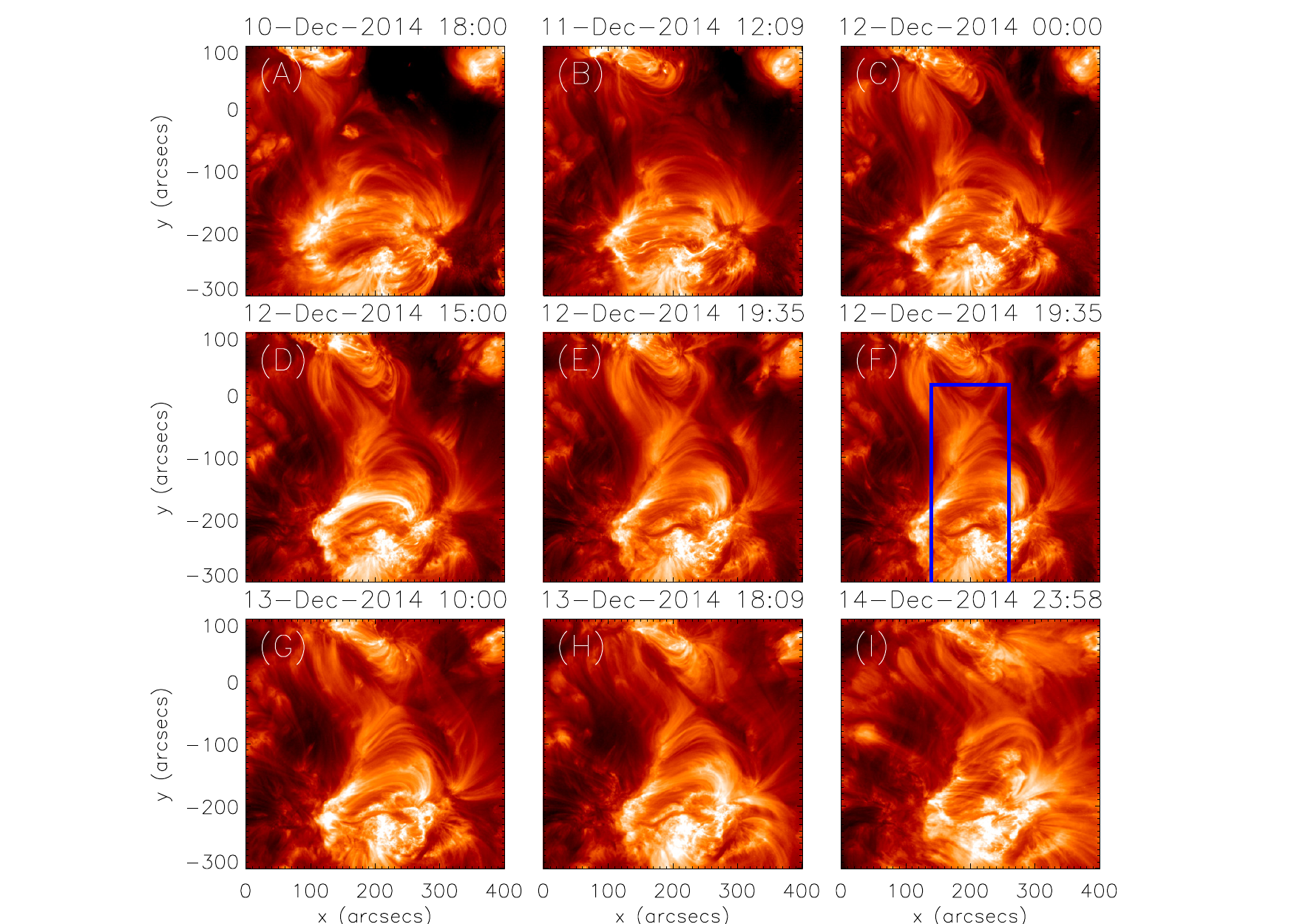}
\caption{Formation and evolution of the region showing TELs (corresponding to the FOV shown in panel (B) of  Fig.~\ref{context}) as observed in AIA 193~{\AA} images for five days from December 10{--}14, 2014. The over-plotted blue box in panel (F) shows the EIS raster FOV. All the AIA data shown here are differentially rotated to the center time of raster E2.}\label{evolution}
\end{figure*}

The EIS rasters included several spectral lines spread over a broad range of temperatures ($\log\,[T/K]= 5.80 - 6.35$). Table~\ref{eis_lines} lists the spectral lines along with their laboratory wavelengths taken from \cite{BroFS_2008} and peak formation temperatures derived from CHIANTI \citep{DerML_1996, DelDY_2015}. The reference wavelength for Doppler velocities are derived using the relatively cooler lines, \ion{Fe}{viii} and \ion{Si}{vii} for CCD~A and CCD~B, respectively. For this purpose, we have used the averaged spectra for all pixels belonging to rows 50{--}149 (from the bottom) of the EIS raster (highlighted by the white box in panels A and B of Fig.~\ref{context}) because these represent the quiet Sun better. However, we have ignored the bottom 0{--}49 rows due to missing data. We have used the eis\_auto\_fit.pro package\footnote{EIS Software Note No. 16, P. Young, 2015} that rectifies the EIS spectral data by removing the errors due to orbital drift and slit tilt. An uncertainty of $\sim$5~km~s$^{-1}$ is expected on EIS velocity estimates \footnote{EIS Software Note No. 17, Peter Young, April 2013}.

\begin{figure*}[!htbp]
\centering
\includegraphics[trim=0.cm 1.cm 0.cm 0.cm,width=1.0\textwidth]{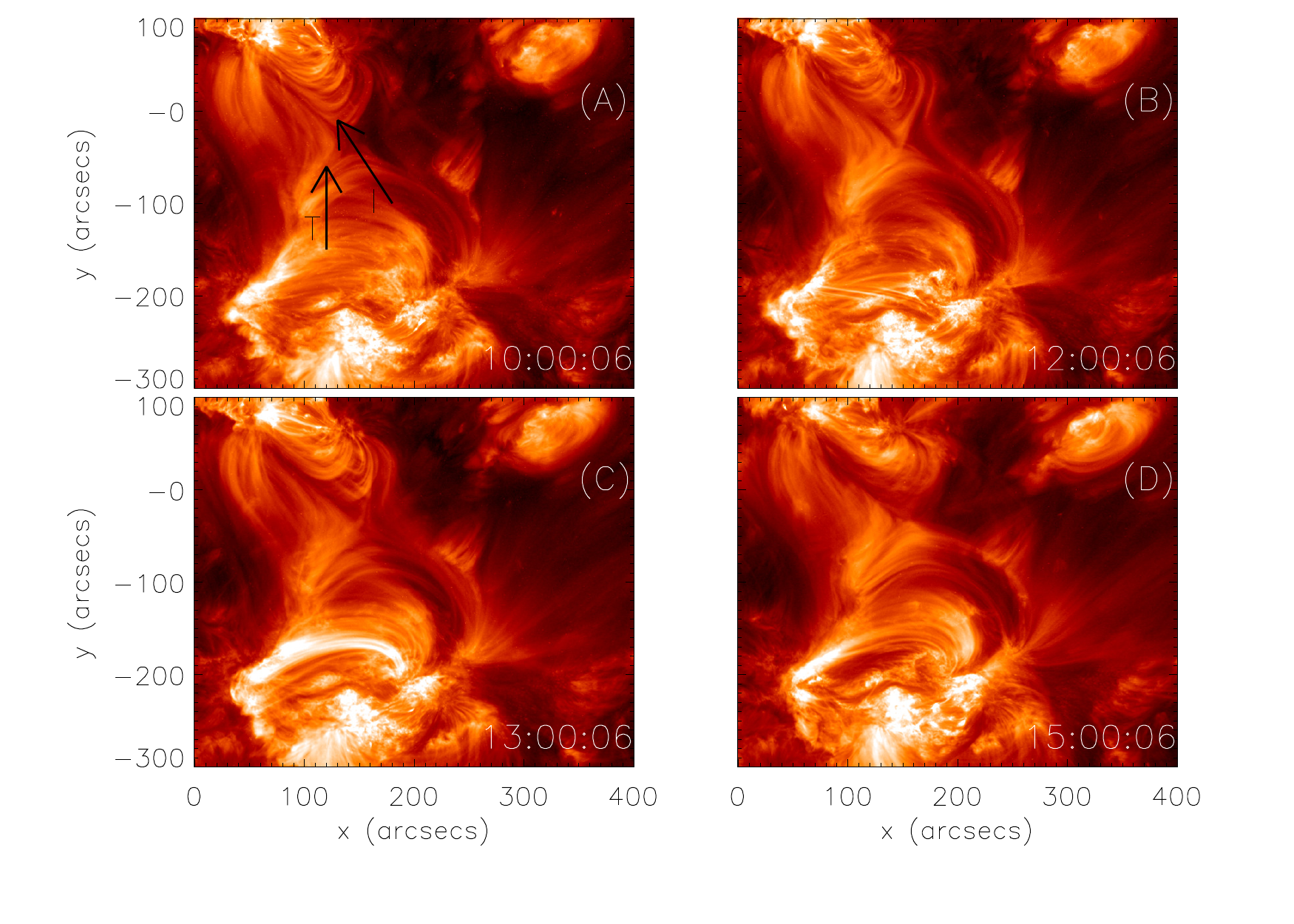}
\caption{Snapshots of events on December 12, 2014 showing formation of TELs and reformation of the loops belonging to individual ARs as a consequence of successive reconnections, as observed by AIA 193~{\AA} channel. For more details, please see the animation  tel12dec.gif.}\label{reform}
\end{figure*}

We note that there are four line pairs, $viz.,$ \ion{Fe}{xi}, \ion{Fe}{xii}, \ion{Fe}{xiii} and \ion{Fe}{xiv} (labelled with a, b in Table~\ref{eis_lines}), which are suitable to estimate the electron densities at four different temperatures \citep[see e.g.,][and reference therein]{YouZM_2007, TriMDY_2010}. For \ion{Fe}{xii}, the spectral line identified at 186.88~{\AA} has two components at 186.854~{\AA} and 186.887~{\AA}. Moreover, \ion{Fe}{xii} 195.119~{\AA} line has a self-blend at 195.179~{\AA}, which is significant ($>5\%$) only at $\log\,N_{e} > $ 9.5 cm$^{-3}$ \citep{YouWH_2009}. We have considered both the lines for our study. The \ion{Fe}{xiii} 203.8~{\AA} line, too, is a combination of 203.79 and 203.82~{\AA} lines, with a component of \ion{Fe}{xii} at 203.72~{\AA}. Therefore, the total intensity under the \ion{Fe}{xiii} line at 203.8~{\AA} is the total of 203.79+203.82~{\AA} intensities. Hence we have fitted two Gaussians to separate out \ion{Fe}{xii} and \ion{Fe}{xiii} lines. The two \ion{Fe}{xiv} lines are not blended and have been fitted with single Gaussians.

\section{Data Analysis and Results}\label{res}
\subsection{Formation and Evolution of TELs}\label{long}

Fig.~\ref{evolution} displays the evolution of the two ARs between December 10{--}14, 2014 as observed with AIA using 193~{\AA} filter. The AR \textsl{12230} in the southern hemisphere first appeared on the visible side of the solar disk on 5th December 2014\footnote{as inferred from www.solarmonitor.org}. However, it was identified as an AR and given its number on 7th December 2014. The other AR \textsl{12234} located in the northern hemisphere was first detected on 10th December 2014. To show the evolution of the structures better, note that we have differentially rotated all the images displayed here at 19:35:30~UT on 12th December 2014 that corresponds to the center time of the EIS raster E2. Fig.~\ref{evolution} demonstrates that as early as on 10th December, the two ARs have individual loop structures along with a slight hint of faint TELs, connecting the two ARs. These TELs become prominent on 12th December (panel C), along with an X-shaped structure that is observed between x $=$ [140$\arcsec$, 260$\arcsec$] and y $=$ [-80$\arcsec$, -20$\arcsec$] (see panels D, E and F). The X-shaped structure exists through 12th December. By 13th December, the ARs drift across the disk, towards the western limb, with gradual disorientation of the loops. The very well-defined morphology of the TELs are no longer visible on or after 14th December. 

An animation \textit{tel12dec.gif} created with data taken at a cadence of 30 minutes using the AIA 193~{\AA} channel images on 12th December, shows a continuous interaction amongst the loops belonging to the individual ARs as well as the TELs. On the one hand, we identify epochs when loops belonging to individual ARs interact and form TELs (e.g., refer to the frames at 13:00 hours in the animation). On the other hand, we also identify the interaction among the TELs leading to the reformation of loops belonging to their parent ARs. To highlight this, we show in panel A of Fig.~\ref{reform} loops belonging to individual ARs (marked as `I'), along with some TELs on the left of the FOV (marked as `T'). In panel B, we see the `I' loops coming close enough so as to merge and form new TELs on the right of the X-region. Similarly, we see these TELs interacting amongst themselves in panel C. Finally, in panel D, we see such TELs have been annihilated to form more loops belonging to the two ARs. 


\subsection{Loop dynamics}\label{dyna}

\begin{figure*}[!htbp]
\centering
\includegraphics[width=1.0\textwidth]{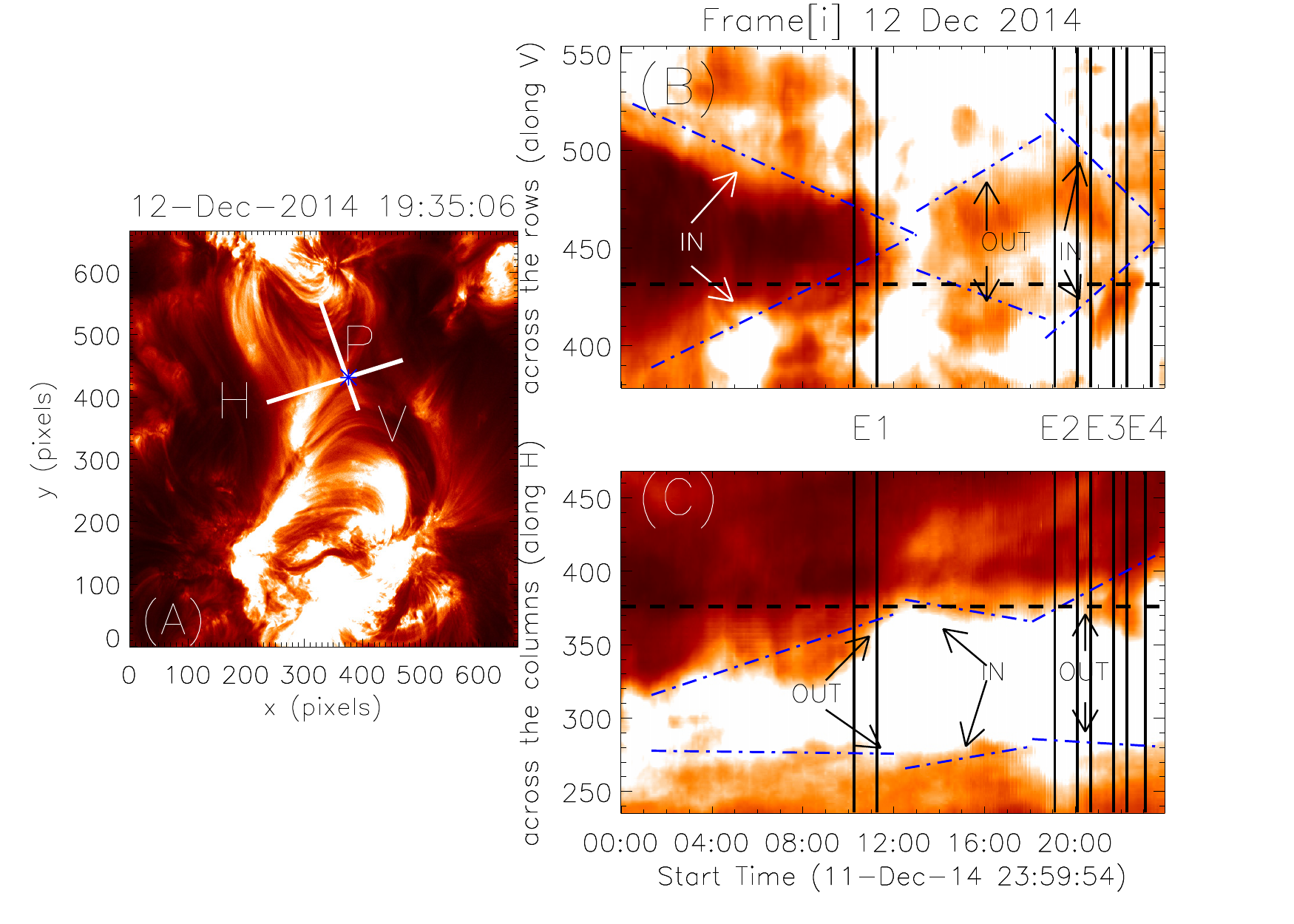} 
\caption{Panel (A): The TELs seen in AIA 193~{\AA} channel in pixel coordinates at a given instant belonging to EIS raster E2. The white lines are the slits, marked as `H' and `V' and the point `P' (`*', in blue) denotes the intersection. Panels (B) and (C): The intensity fluctuations along the slits `V' and `H' respectively in the xt plots. The over-plotted solid black lines indicate the EIS raster durations, with numbers `E1', `E2', `E3' and `E4' indicating the respective raster number. The dashed-dotted blue lines track the displacement of plasma with time. The black dashed line shows the position of the point `P' in pixel coordinates.}\label{xt}
\end{figure*}

To further understand the dynamics of the loops, we created xt-plots along two artificial 10-pixels wide slits named `V' and `H' (see panel A in Fig.~\ref{xt}). For this purpose, we have used the data taken on December 12, 2014, with AIA~193{\AA} at a cadence of 1 minute. Panels B and C of Fig.~\ref{xt} show the xt-plots corresponding to the slits V and H, respectively. The horizontal black dashed lines trace the locus of point $P$ labelled in panel A whereas the solid vertical black lines locate the four EIS raster phases, marked as E1, E2, E3 and E4. 

The xt-plot corresponding to the slit V (panel B) demonstrates the inward movement of the structures from the very beginning and appear to merge soon after the raster E1, approximately between 12:00 and 13:00 UT. Such inward motion in the xt-plot along the slit V suggests the evolution of loops belonging to individual active regions and moving closer to each other. During the same interval, the xt-plot corresponding to slit H (panel C) reveals hints of outward motion, though not as systematic as the case for the inward movement. The outward movement in the xt-plot for slit H suggests the structures are moving away. Combining the two xt-plots, we infer that loops belonging to the individual active regions are evolving. Their tips are coming closer to each other, leading to magnetic reconnection and formation of the TELs. We have highlighted the inward and outward motions with blue slanted (dashed-doted) lines in panels B and C.

After 13:00 UT, the motions observed in the two xt-plots reverse direction till about 19:00 UT. The reversing of movements suggests that during this period, the TELs are evolving and moving towards each other. During this time, the TELs reconnect and lead to the reformation of the loops belonging to individual active regions. Beyond 19:00~UT on December 12, 2014, the motion is similar to that is observed during the first phase. 

We have used the over-plotted blue dashed-dotted lines in panel B (by eye estimation) to obtain the inflow speeds during the first and third phases. The inflow speeds are 1 and 3~km~s$^{-1}$, whereas the outflow speed for the second phase is 1.5~km~s$^{-1}$. Due to ambiguous patterns observed in the xt-plot shown in panel C, we did not measure speeds along the slit H. Similar values of inflows and outflow speeds were also reported by \cite{YokAM_2001} in an observation related to a flare.

\subsection{Identification of the reconnection point with AIA images}\label{iden_aia}

\begin{figure*}[!htbp]
\centering
\includegraphics[width=1.0\textwidth]{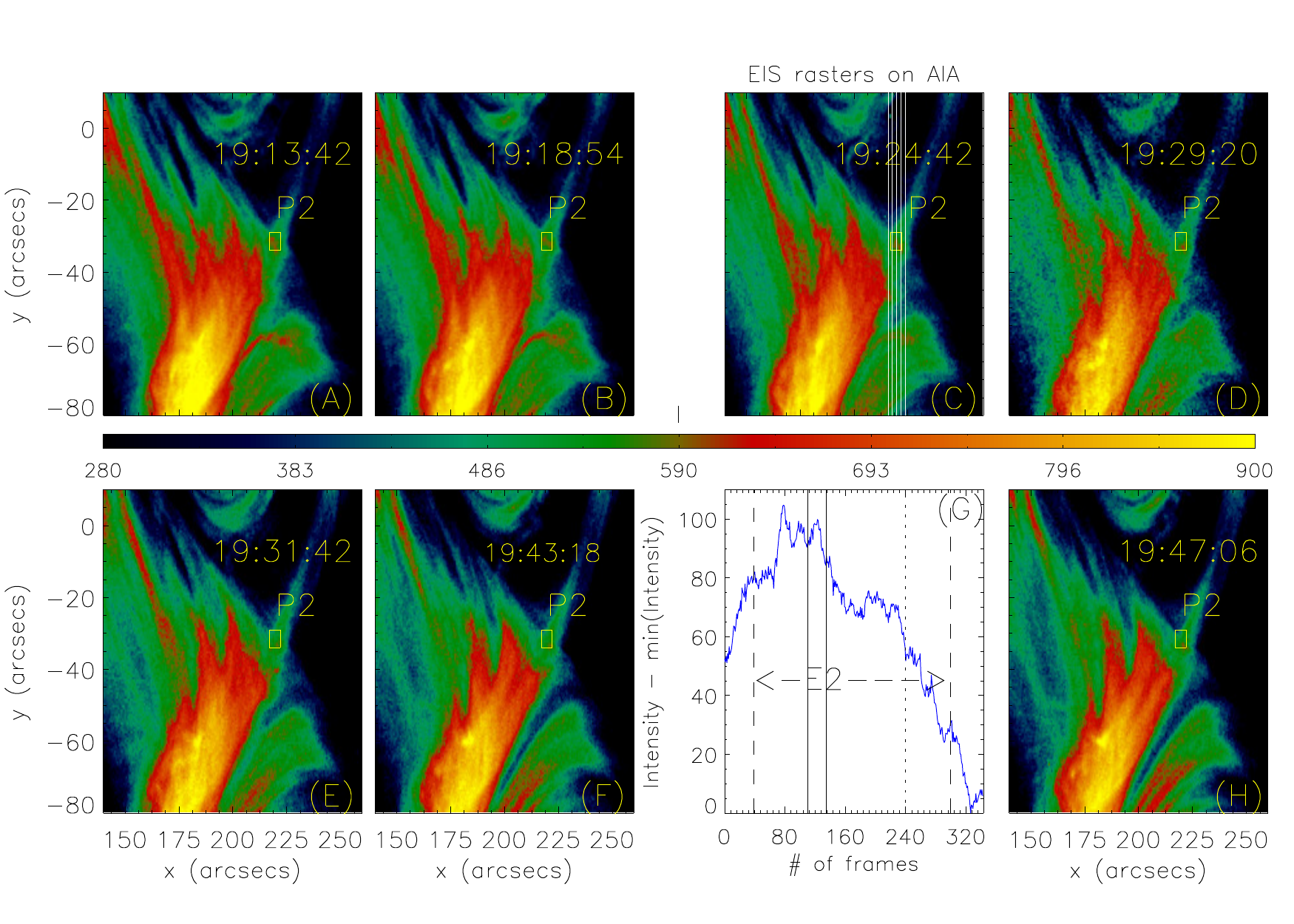}
\caption{Panels (A){--}(F) \& (H): AIA 193~{\AA} intensity images at different instants, roughly coinciding with the EIS observation period for E2. `P2' (highlighted by the yellow box) denotes the location for recurrent brightenings. The white vertical lines in panel (C) show that four EIS exposures which cover the location of this brightening. Panel (G): The corresponding light curve in `P2', between 18:55:07{--}20:14:55 UT at a cadence of 12 seconds. The exact duration of E2 is marked by the dashed vertical black lines. The black bold lines indicate the increase in intensity in `P2' corresponding to the EIS exposures shown with white lines in panel C.}\label{dots}
\end{figure*}

As mentioned earlier, the X-region was rastered with EIS four times on 12th December. Since the raster takes time to scan through the FOV, we identify a region that was simultaneously observed by AIA and EIS slits. For this purpose, we plot in Fig.~\ref{dots} the AIA intensity maps in 193~{\AA} channel corresponding to `E2' (panels A{--}F and H). However, the ROI plotted in these maps is between y=[-80$\arcsec$, 10$\arcsec$]) so as to highlight the dynamics/ features in the neighbourhood of the X-region. We identify intermittent brightenings occurring between x=[217$\arcsec$, 222$\arcsec$] and y=[-34$\arcsec$, -29$\arcsec$], which is henceforth identified as `P2' (marked with the yellow box). Coincidentally, this region is also observed by four EIS raster exposures (highlighted by the white vertical lines in panel C).  

In panel G (just beneath panel C), we plot the average light curve in this box with a cadence of 12 seconds between 18:55:07{--}20:14:55~UT. The bold black vertical lines represent the four exposures of E2 raster, each 2$\arcsec$ wide, corresponding to the white lines marked in panel C. In addition, the exact duration of E2 is marked by the dashed black vertical lines in panel G. From the light curve, we see that three intensity peaks occur in quick succession at P2. This indicates dynamic events occurring and are also captured by simultaneous EIS raster observations. This gives us the opportunity for measuring the plasma parameters in the P2 region. Similarly, for E4 observations, there is a partial overlap of EIS observations with the region having repeated intensity enhancement. However, no such spatial overlap of EIS exposures with those of homologous intensity enhancement is observed for E1 and E3. 

\begin{figure*}[!htbp]
\centering
\includegraphics[width=1.0\textwidth]{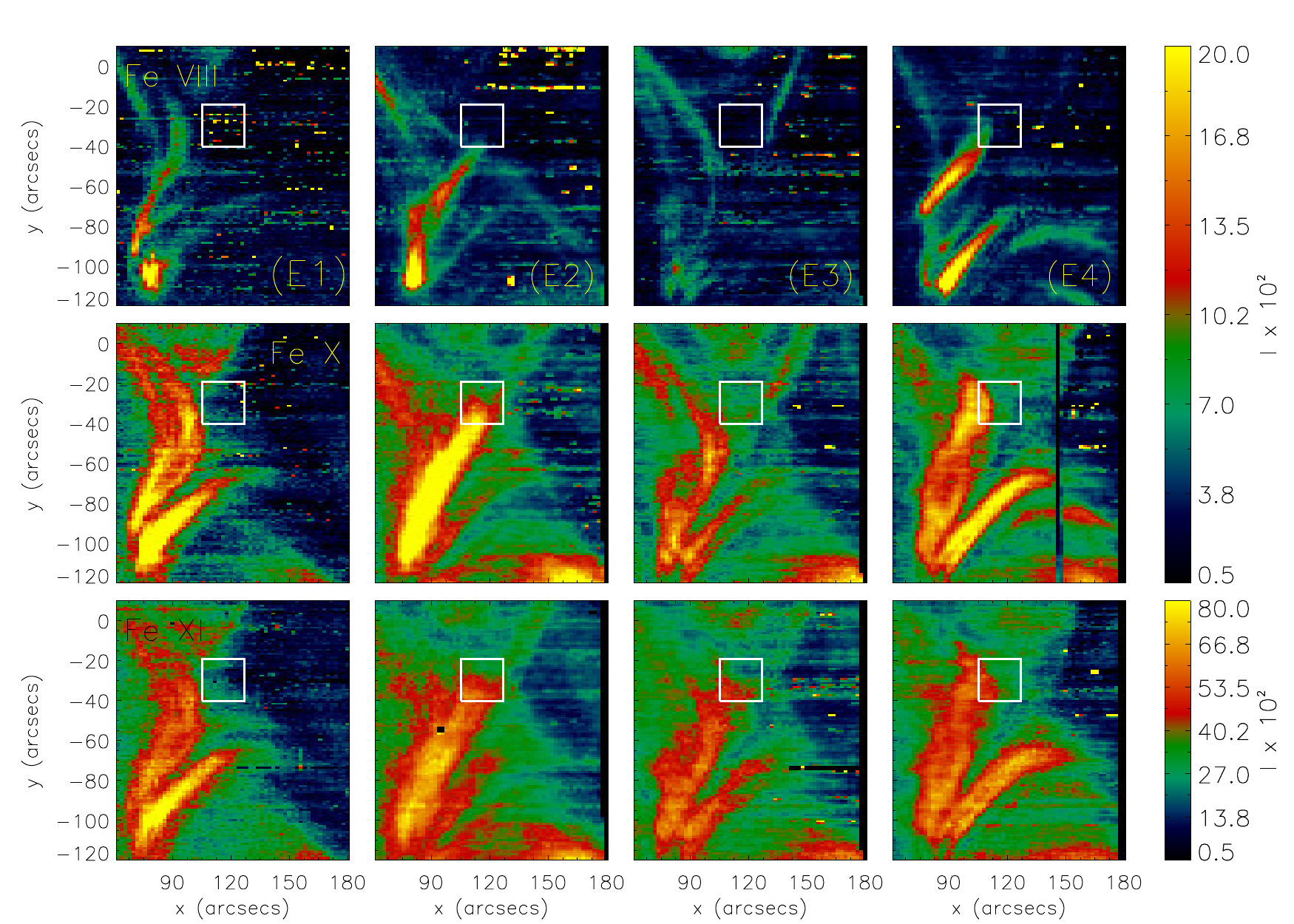}
\caption{The intensities in EIS \ion{Fe}{viii}, \ion{Fe}{x} and \ion{Fe}{xi} lines for the four events. The FOV of E2, E3 and E4 have been rotated to that of E1. The white box shows the reconnection region. The raster number on the top panel of each column is true for all the spectral lines whereas the each row corresponds to the same spectral line as mentioned in the leftmost panel of each row. }\label{eis_int_1}
\end{figure*}

\subsection{Plasma Diagnostics with EIS}\label{plas}

\begin{figure*}[!htbp]
\centering
\includegraphics[width=1.0\textwidth]{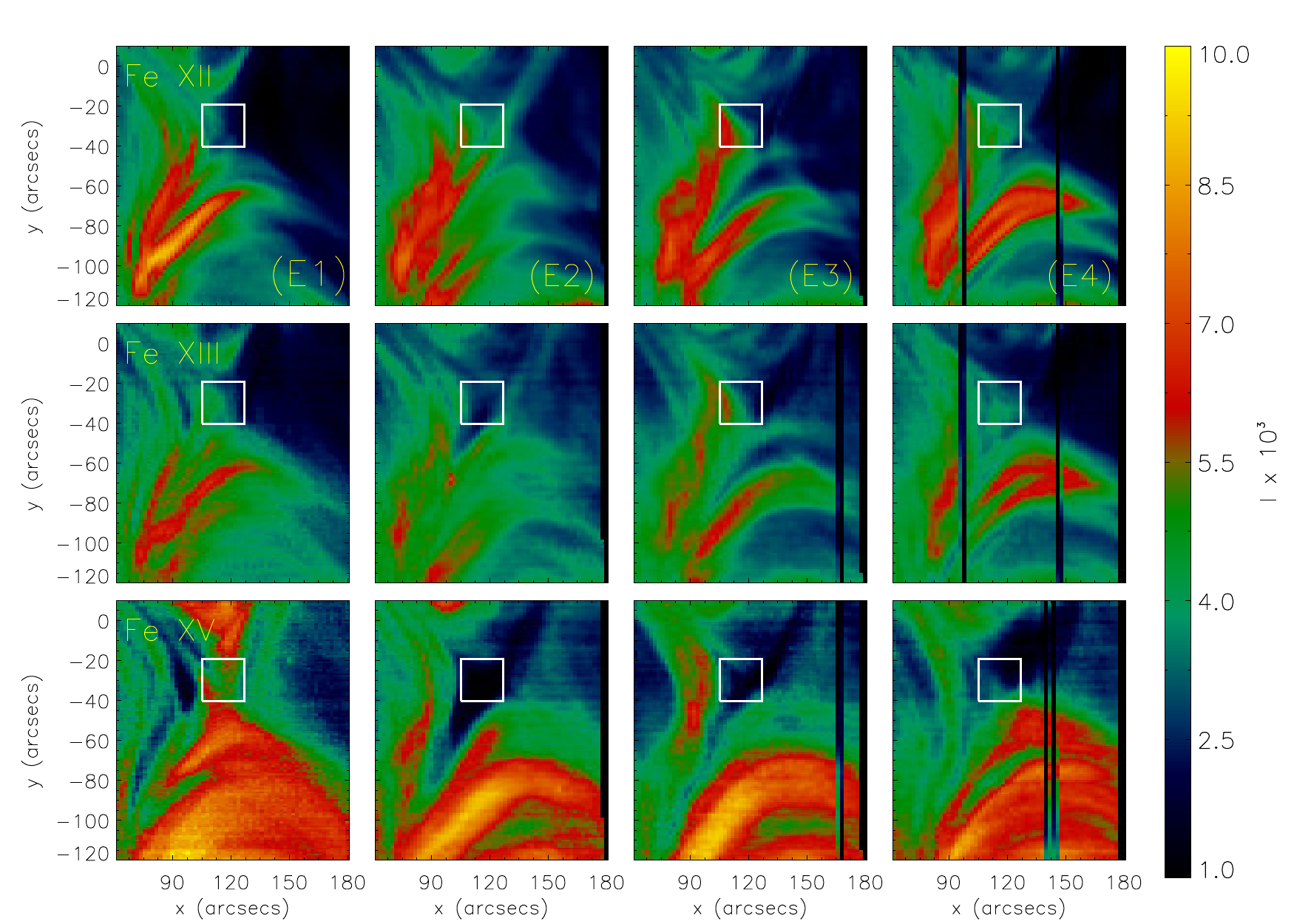}
\caption{Same as Fig.~\ref{eis_int_1} but for \ion{Fe}{xii}, \ion{Fe}{xiii} and \ion{Fe}{xv} lines.}\label{eis_int_2}
\end{figure*}

As stated earlier and shown in Fig.~\ref{context}, the EIS spectrometer observed the X-shaped region four times on 12th December 2014, in raster mode. These are denoted as E1, E2, E3 and E4. We plot the EIS intensity maps obtained in different ionisation states of Iron from \ion{Fe}{viii} to \ion{Fe}{xv} in Figs.~\ref{eis_int_1} \& \ref{eis_int_2} (as labelled). In these maps, we emphasize the region between y=[-120$\arcsec$, 10$\arcsec$], so as to show the morphology of the loops near the X-region as well as the loops beneath and above it, in detail. The four columns in these two figures correspond to the four different EIS rasters. We have differentially rotated the FOVs of E2, E3 and E4 to the raster start time of E1, to facilitate the time evolution study of the same region. The noticeable rightmost black stripes in E2, E3 and E4 are due to this rotation. We locate the X-shaped region in all the intensity maps with a blue box, that encloses the point `P' shown in panel A of Fig.~\ref{xt}.

The intensity maps obtained at cooler temperatures, such as in \ion{Fe}{viii} and \ion{Fe}{x}, show a number of distinct loops criss-crossing each other (top and middle rows in Fig.~\ref{eis_int_1}). Similar structures are also seen in the intensity images obtained in \ion{Fe}{xi} (bottom row of Fig.~\ref{eis_int_1}) and \ion{Fe}{xii} (top row of Fig.~\ref{eis_int_2}) but with a significant amount of diffuse emission, similar to those observed by, for $e.g.$, \cite{TriMD_2009, SubTK_2014}. In the intensity images obtained at higher temperatures (exceeding $\log\, T[K]=$ 6.20), we notice a decrease in the intensity in the X-shaped region. The region appears fully evacuated in the intensity maps of \ion{Fe}{xiv} and \ion{Fe}{xv} for all the three rasters but E1 (bottom row of Fig.~\ref{eis_int_2}). It is worth noticing that there is an abundance of bright loops in the \ion{Fe}{xv} intensity maps beneath the X-region.

\begin{figure*}[!htbp]
\centering
\includegraphics[trim=45 5 100 6,clip,width=0.48\textwidth]{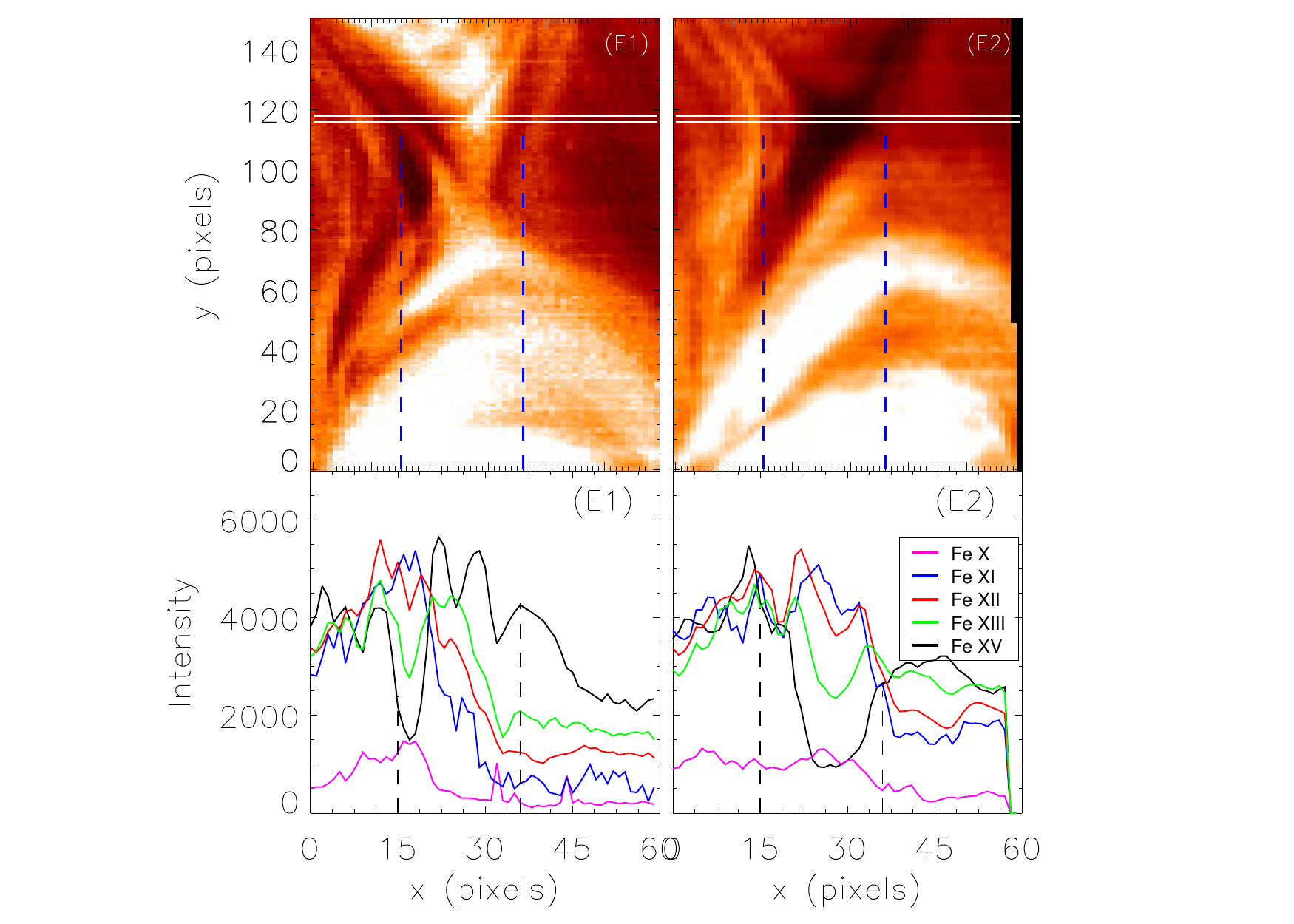}
\includegraphics[trim=45 5 100 6,clip,width=0.48\textwidth]{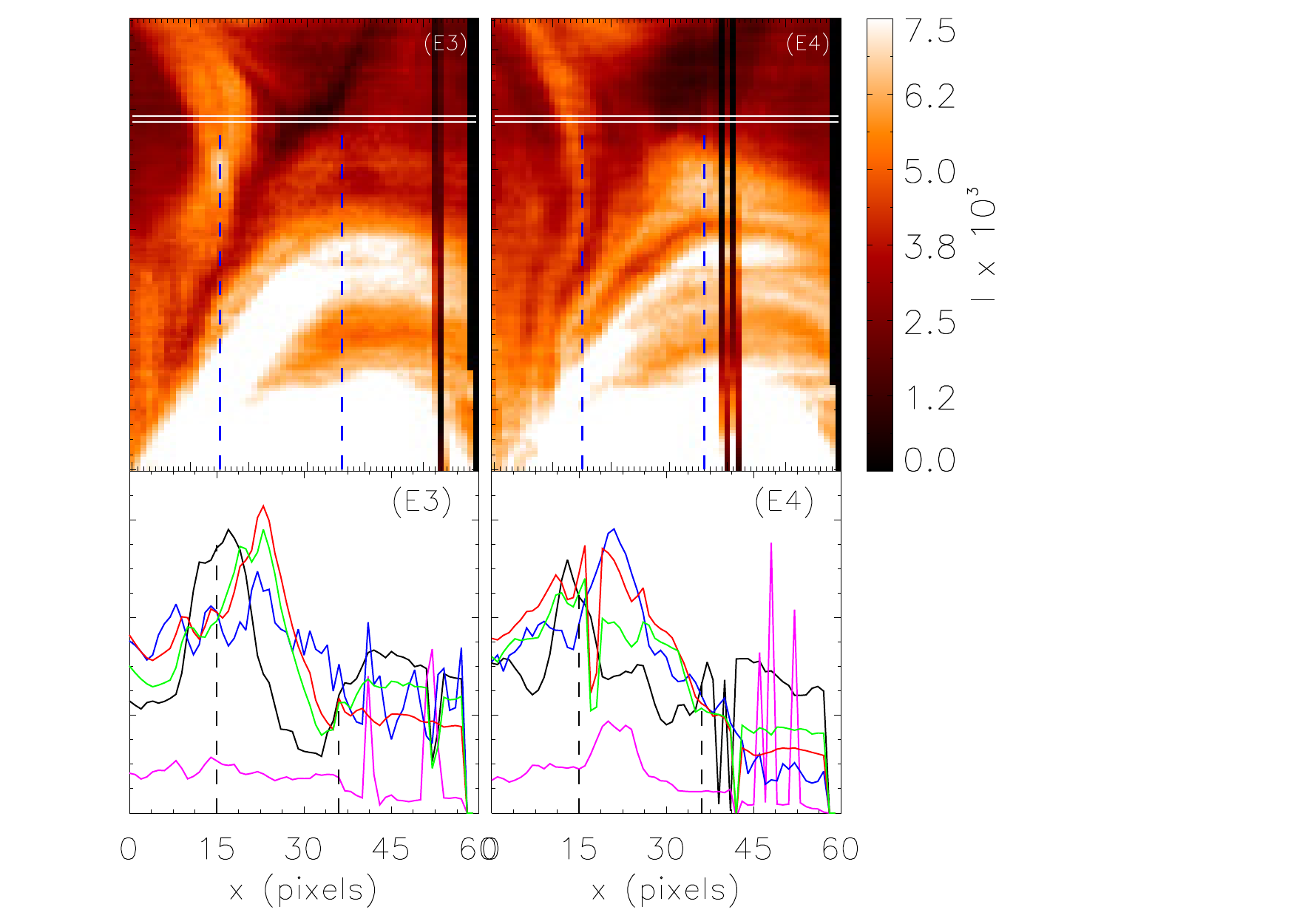}
\caption{Top panels: EIS intensity maps obtained in \ion{Fe}{xv}. The horizontal white stripe is the region for which the averaged light curve is obtained. Bottom panels: Average light curves within the white stripes in spectral lines as labelled. The vertical black dashed lines in the bottom panel(s) correspond to the same pixel position as indicated by the vertical blue dashed lines in the top panel(s), highlighting the intensity variation in the X-region.}\label{lc}
\end{figure*}

To illustrate this better, we obtain light curves in five spectral lines of Iron for all four EIS rasters and plot them in Fig.~\ref{lc} as labelled. The top panels are the intensity maps in \ion{Fe}{xv} and bottom panels are the light curves. We obtain the light curves by averaging the intensities within the two horizontal white lines overplotted in the top panels. We have drawn the vertical blue and black lines in the top and bottom panels, respectively, to highlight the extent (in x-direction) of the X-shaped region and the corresponding extent in the respective light curves. 

The light curves (shown in the bottom panels of Fig.~\ref{lc}) reveal that the X-shaped region is dimmer in all the spectral lines in all rasters, except E1, where \ion{Fe}{xv} is brighter than other lines. However, it shows a decrease in intensity where \ion{Fe}{x} shows an enhancement. Similar nature of light curves is noted for \ion{Fe}{xiv} (not shown here), albeit it is a weak line and has minor fluctuations in intensity as a function of position.


\subsubsection{Density and Temperature diagnostics} 

\begin{figure*}[!htbp]
\centering
\includegraphics[trim=0.cm 0.cm 0.cm 1.cm,width=1.0\textwidth]{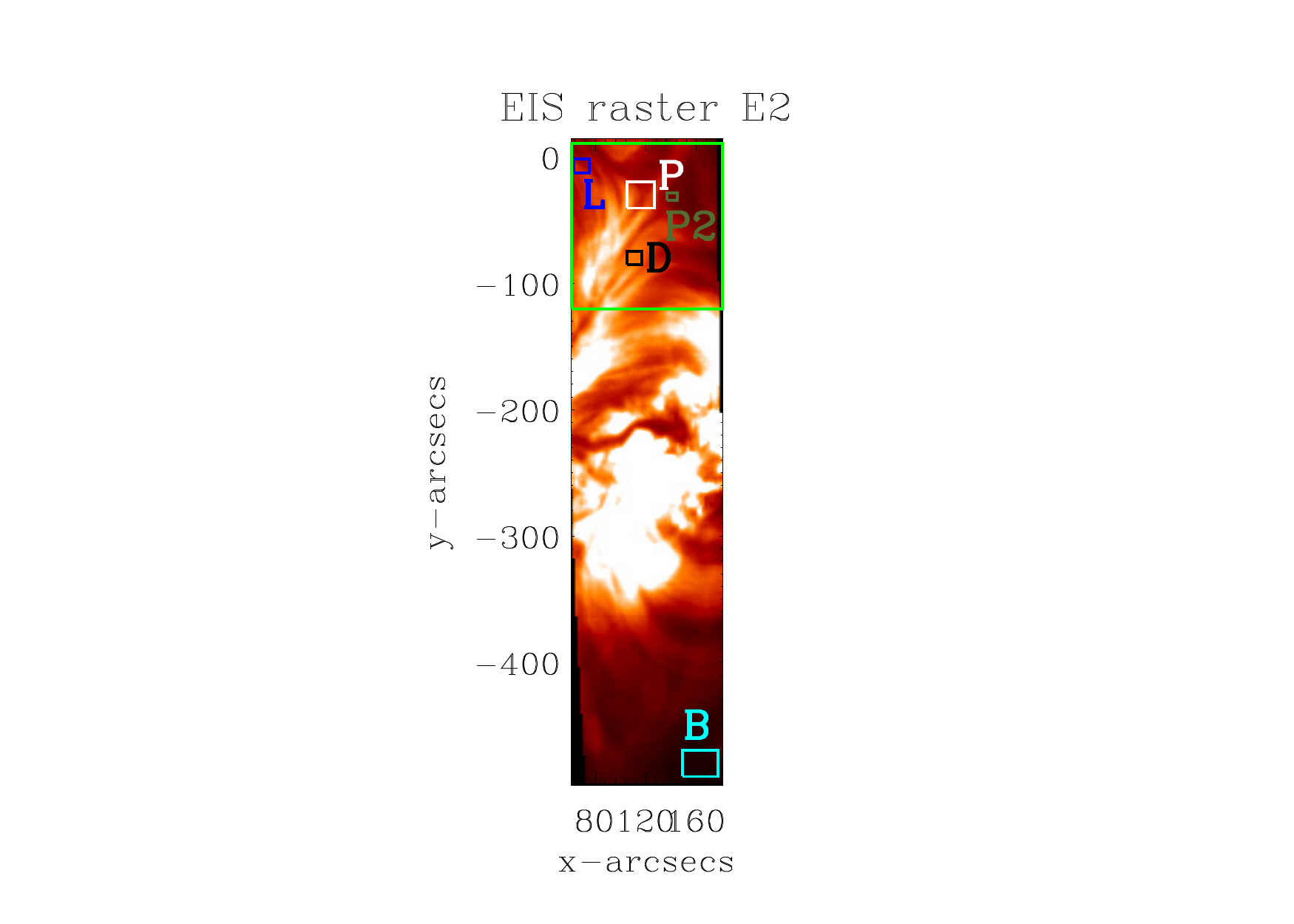}
\caption{Identification of different regions in the FOV for the TELs (`L', in blue), X-shaped region (`P', in white) and the hot loops beneath (`D', in black) on the background intensity map of \ion{Fe}{xii} 195~{\AA} at E2. `P2' (in olive) is the region which undergoes several brightenings at regular intervals within the EIS observation period E2 and has four EIS exposures, coinciding exactly with the position of the brightening. The box identified with `B' marks the region considered for background/ foreground assessment. The large green box indicates the ROI of the study.}\label{tels_loc}
\end{figure*}

Plasma diagnostics are conducted in several regions within the EIS raster FOV, to discern the properties in different types of loop structures captured in the same. The average electron densities and temperatures have been estimated at the X-region along with those in the adjacent loops for all four rasters. These are denoted in Fig.~\ref{tels_loc} by `P' (X-region), `L' (TELs) and `D' (hot loops beneath the X-region) on the \ion{Fe}{xii} 195~{\AA} intensity map of E2, rotated with respect to E1. In addition, for E2, we have the fourth region of interest P2 which shows intermittent intensity enhancements, as shown in panel G of Fig.~\ref{dots}.


\paragraph{Assessment of background/foreground \newline} \label{back} 

Estimate of background/ foreground emission plays an important role in measuring plasma electron densities and temperatures \citep{DelM_2003, TriKM_2011}. For this, we identify a region marked as `B' in Fig.~\ref{tels_loc}. The average intensity in `B' is assumed to be the background/ foreground intensity. The green box in Fig.~\ref{tels_loc} is the zoomed-in ROI and correspond to the ROI that is shown in Figs.~\ref{eis_int_1} and ~\ref{eis_int_2}. 

\begin{figure*}[!htbp]
\centering
\includegraphics[trim=1.cm 3.cm 0.cm 3.cm,width=1.0\textwidth]{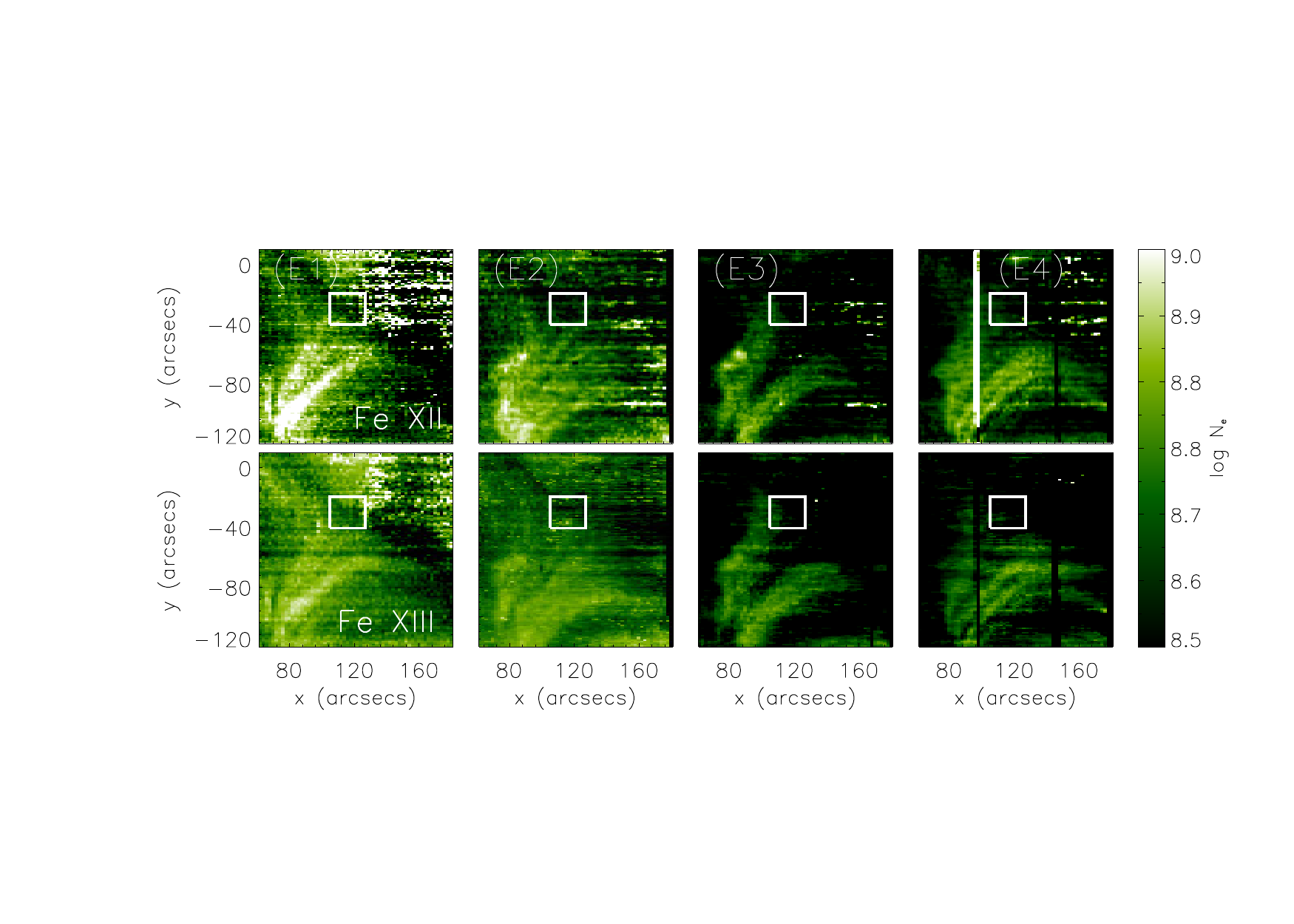}
\caption{Density maps using EIS lines \ion{Fe}{xii} ($\log\,T[K]=$ 6.20) and \ion{Fe}{xiii} ($\log\,T[K]=$ 6.25). The raster number on the top panel of each column is true for all the spectral lines whereas the each row corresponds to the same spectral line as mentioned in the leftmost panel of each row. The white box in each panel shows the X-region. Backfround/ foreground emission has been accounted for.}\label{density_maps}
\end{figure*}

\begin{figure*}[!htbp]
\centering
\includegraphics[width=0.49\textwidth]{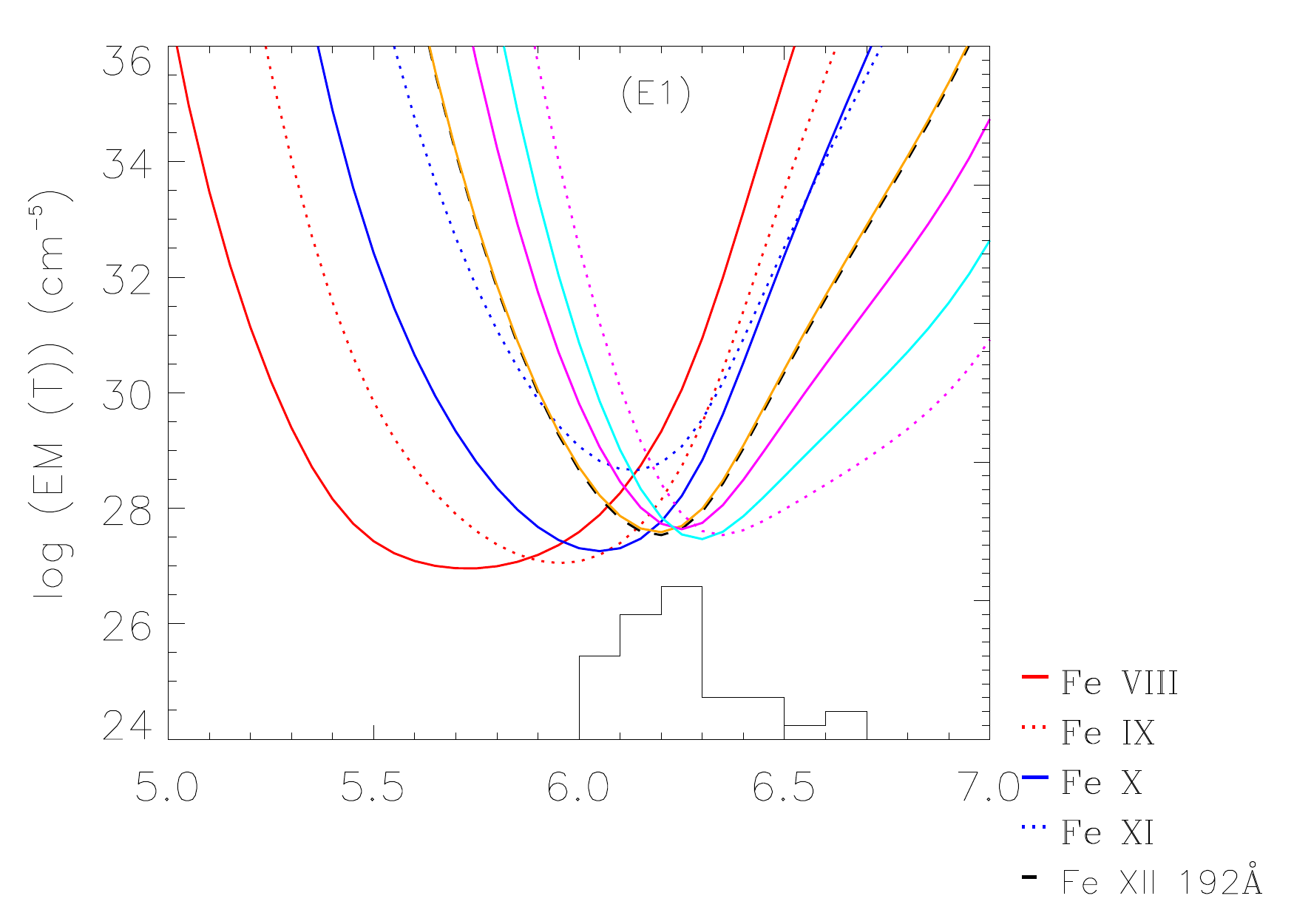}
\includegraphics[width=0.49\textwidth]{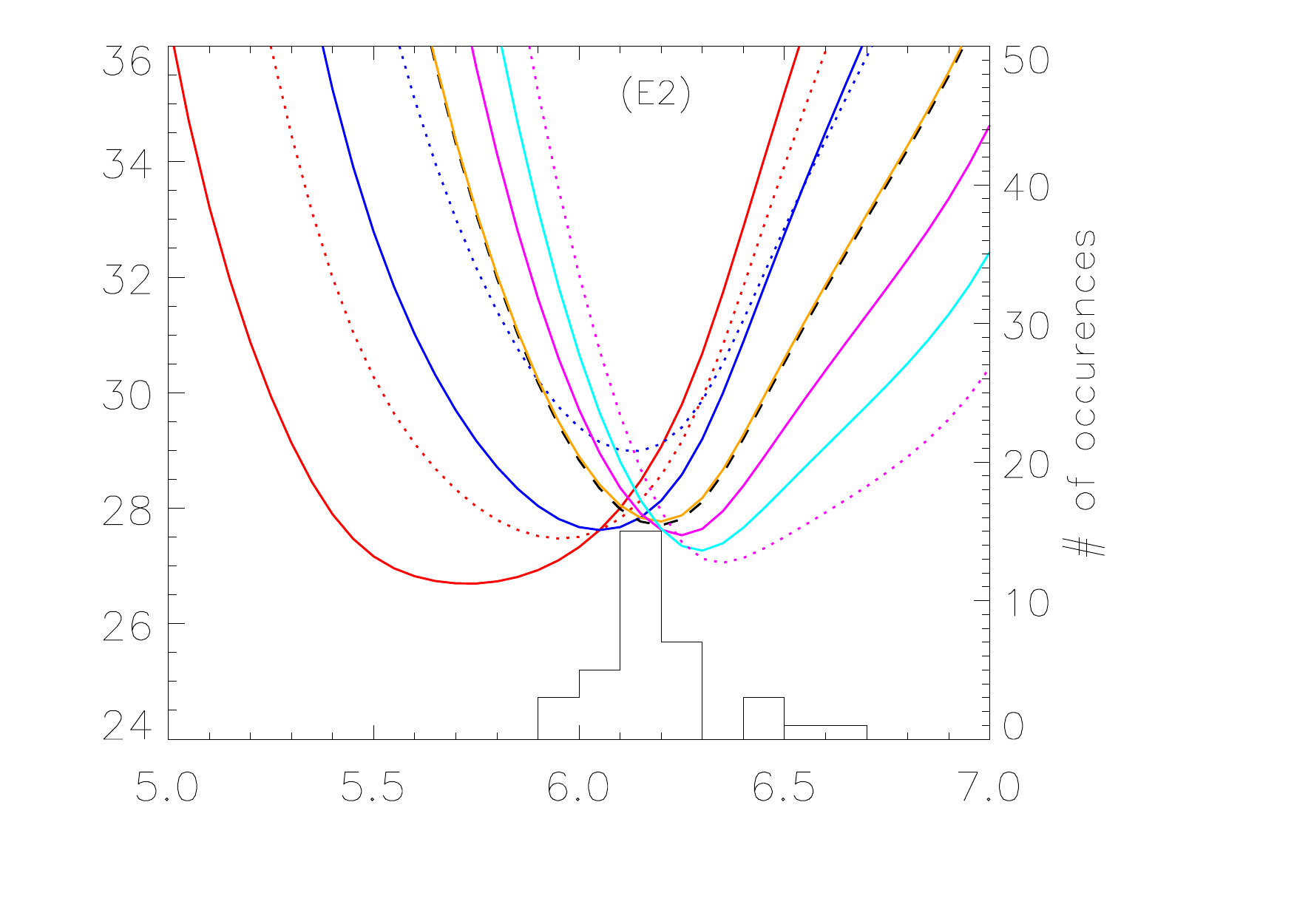}
\includegraphics[width=0.49\textwidth]{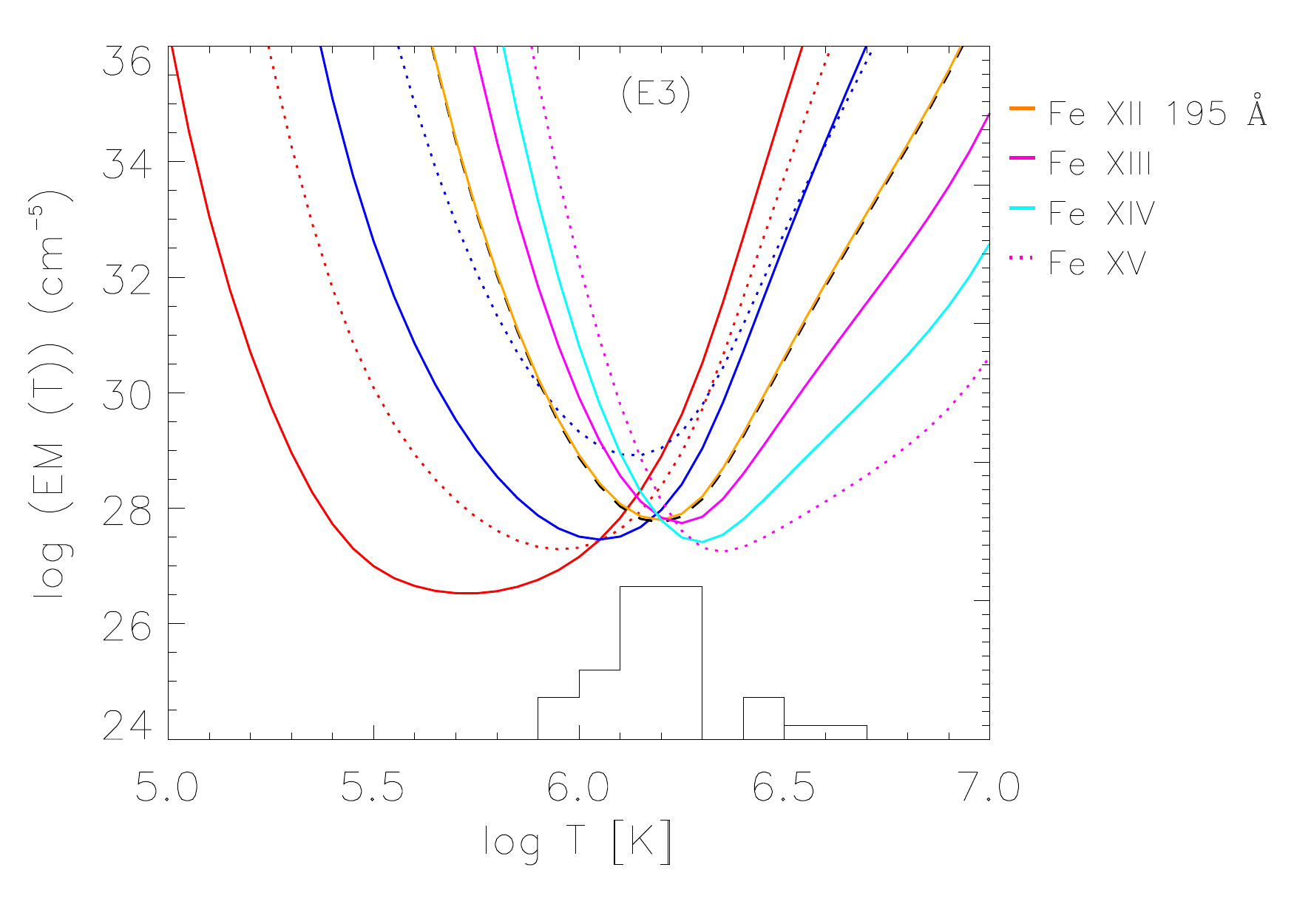}
\includegraphics[width=0.49\textwidth]{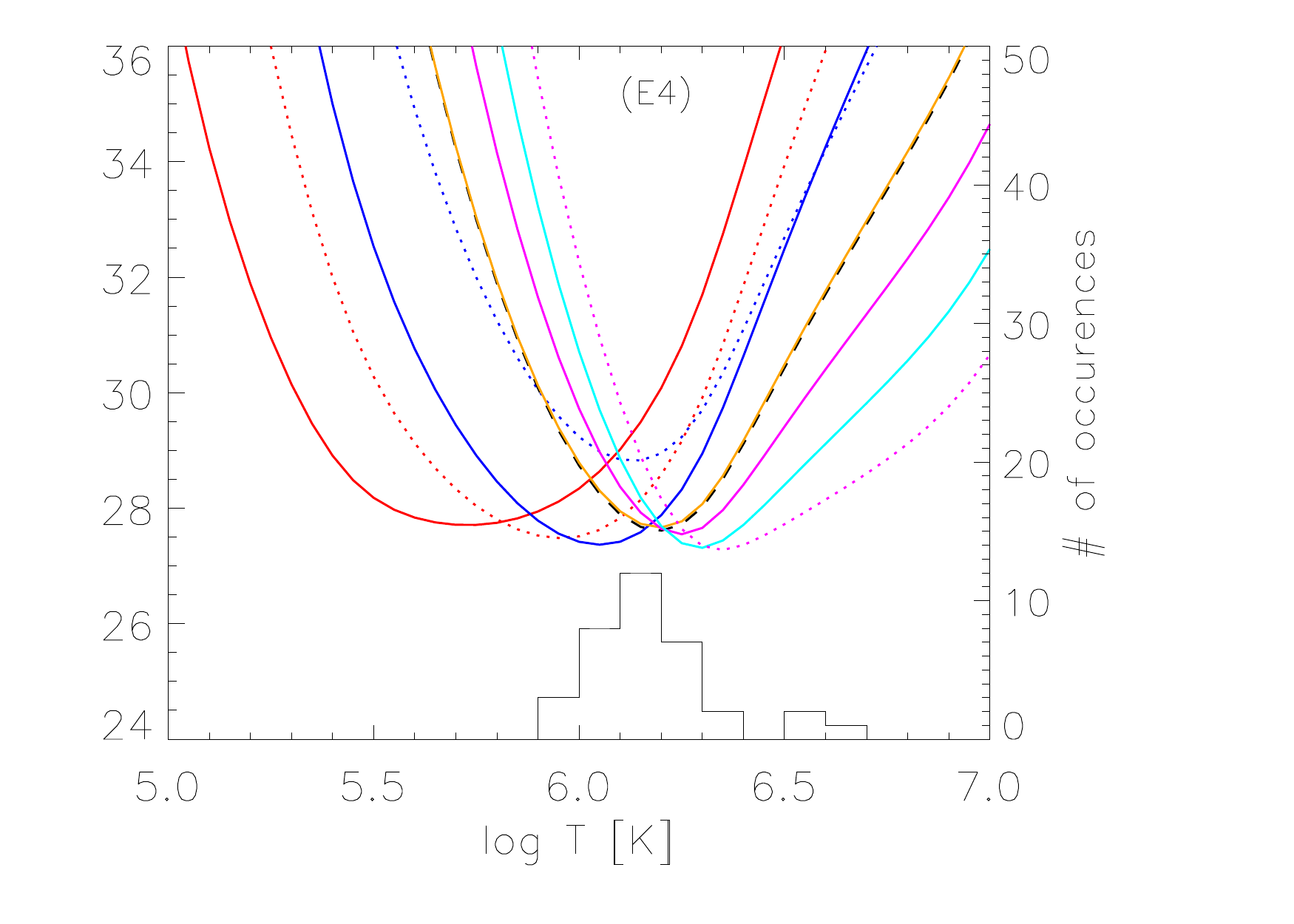}
\caption{Emission Measure (EM) loci curves for the four EIS rasters for the X-region as shown in Fig.~\ref{density_maps}. Also plotted are the histograms representing number of crossings within a given temperature bin. No background/ foreground emission has been considered.}\label{em_loci}
\end{figure*}

The averaged intensities for locations P, P2, L, D and B are noted in Table~\ref{inten}. We note that the averaged intensities in \ion{Mg}{vii} and \ion{Si}{vii} in `B' region larger than the actual locations which are considered for the estimates. Similar discrepancy is noted for \ion{Fe}{xiv} line. We attribute this to the weak spectral lines and therefore, discard these three lines for any kind of plasma diagnostic studies. In addition, we note that for different locations for different rasters, the average intensities in `B' exceed those in the other three regions. This renders most of the low temperature lines $viz.,$ \ion{Fe}{viii}, \ion{Fe}{ix}, \ion{Fe}{x} and \ion{Fe}{xi} unsuitable for the EM-loci computation. That leaves only a set of four lines \ion{Fe}{xii} 192~{\AA} and 195~{\AA}, \ion{Fe}{xiii} and \ion{Fe}{xv} only. Therefore, we neglect the background/ foreground correction for EM-loci plots for all three (four) regions, `P', `L', and `D' (and `P2' in case of E2). 

\paragraph{Density diagnostics in the X-region \newline} \label{dens} 

Using the average intensities in `B' in Fig.~\ref{tels_loc} as the background/ foreground values, we derive the density maps for \ion{Fe}{xii} and \ion{Fe}{xiii} line pairs (refer to Fig.~\ref{density_maps}) for all four raster periods. Note that these maps show background subtracted densities. The four columns represent the four different rasters as labelled. The over-plotted white-box in each map highlights the X-shaped structure (same location as the blue boxes in Figs.~\ref{eis_int_1}~\&~\ref{eis_int_2}). It is further noted that these boxes enclose the point `P', shown in panel A of Fig.~\ref{xt}. We neglect deriving density maps obtained using \ion{Fe}{xi} and \ion{Fe}{xiv}, since these are noisy. This is because of one of the lines of \ion{Fe}{xi}, i.e. 180.401~{\AA} is at the edge of CCD A where the effective aperture area is small; and \ion{Fe}{xiv} lines have small signal to noise ratios as they are weak. 

The average densities within the X-shaped regions (enclosed by white boxes) for all four rasters using the \ion{Fe}{xii} (similar for \ion{Fe}{xiii}) lines range between $\log\,N_e$ = 8.46 to 8.67~cm$^{-3}$ (also see, Table~\ref{parameters}). To check the consistency of our results, we compared these densities by obtaining the averaged spectra within the white box before taking the ratios. The density estimates were similar to those listed in Table~\ref{parameters}. In addition, the density at the identified brightening observed by EIS slits, $i.e.,$ P2 is $\log\,N_e$ = 8.50 in \ion{Fe}{xii}, which is consistent with the average values in the X-region. 


\paragraph{Temperature diagnostics in the X-region \newline} \label{temp}

\begin{figure*}[!htbp]
\centering
\includegraphics[trim=0.cm 1.cm 0.cm 0.cm,width=1.0\textwidth]{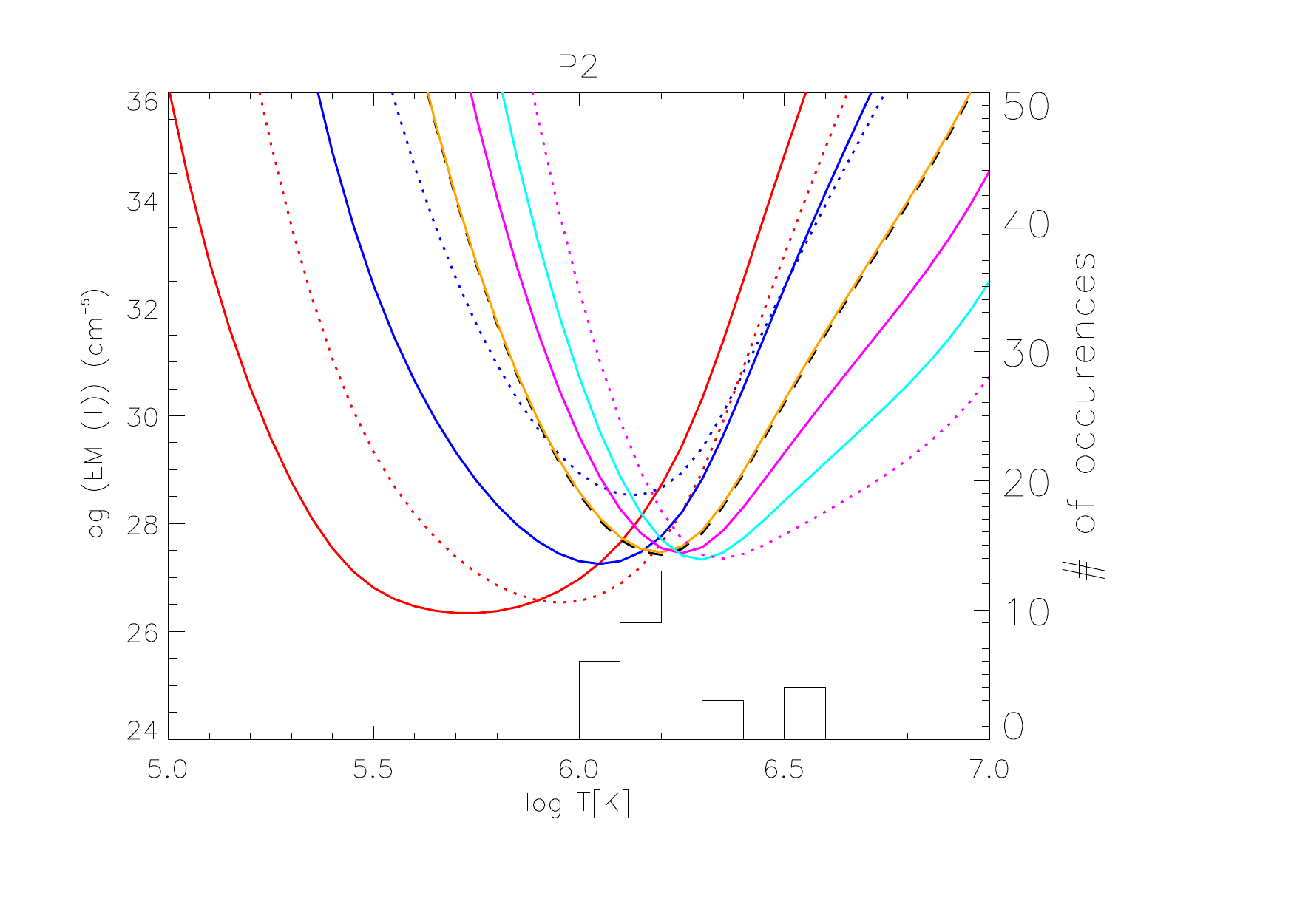}
\caption{Emission Measure (EM) loci curves for the bright point, corresponding to `P2' (in Figs.~\ref{dots} and ~\ref{tels_loc}), derived with spectral line observations during E2. The histograms representing number of crossings within a given temperature bin are plotted. No background/ foreground emission has been accounted for.}\label{em_p2}
\end{figure*}

\begin{figure*}[!htbp]
\centering
\includegraphics[width=0.46\textwidth]{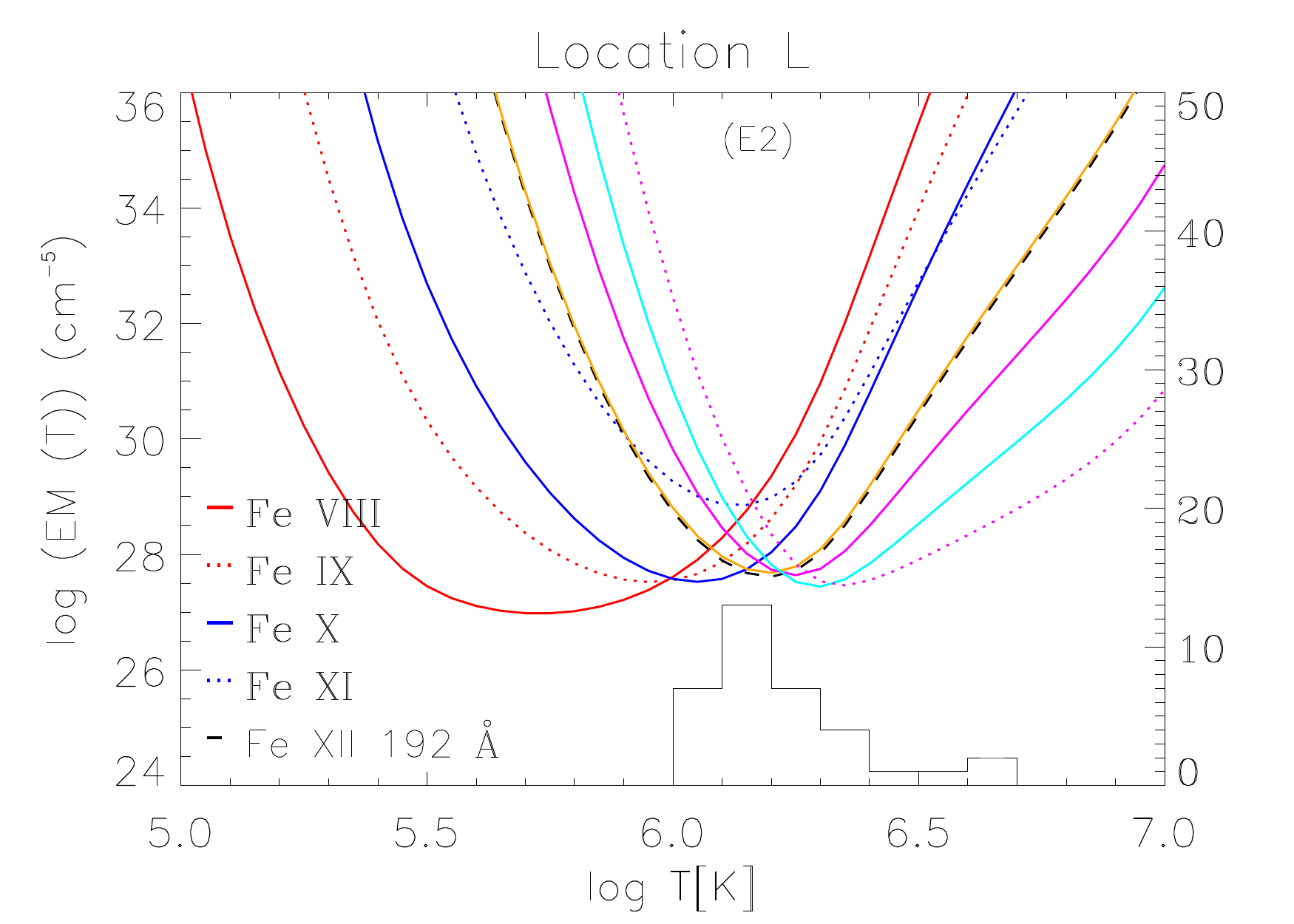}
\includegraphics[width=0.46\textwidth]{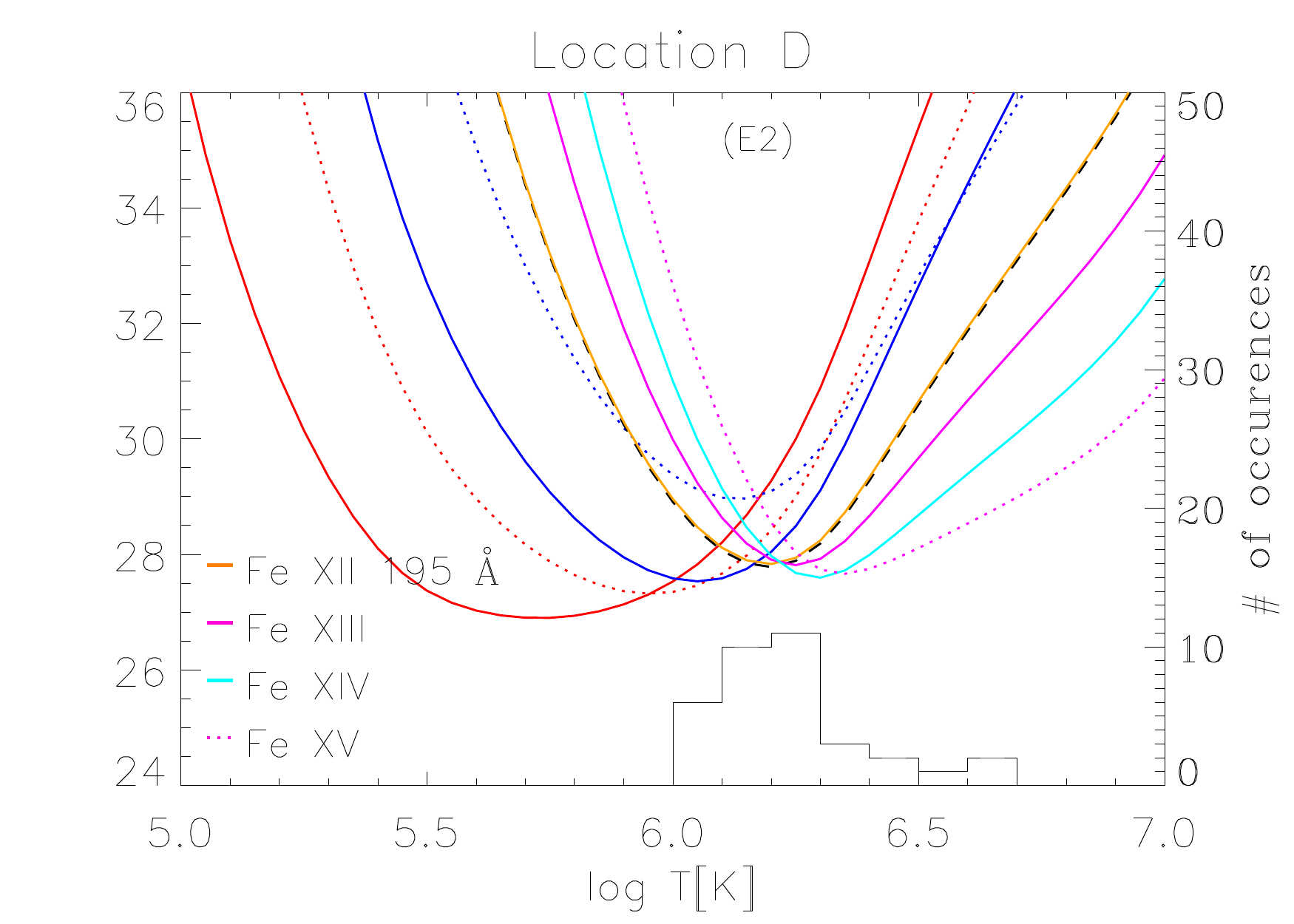}
\caption{Emission Measure (EM) loci curves for the TELs (`L', in left panel) and the hotter loops beneath (`D', in right panel) in E2, as identified in Fig.~\ref{tels_loc}. Also plotted are the histograms representing number of crossings within a given temperature bin. No background/ foreground emission has been considered. }\label{tels_em}
\end{figure*}

We estimate the temperature of the X-shaped region, which is shown by blue boxes in Figs.~\ref{eis_int_1} \& \ref{eis_int_2}, using the Emission Measure (EM) loci curves \citep[][]{JorW_1971, TriMDY_2010}. For this purpose, we first obtain the averaged intensities in the region in all the spectral lines marked with `t' in Table~\ref{eis_lines}. However, we exclude \ion{Si}{vii}, for reasons explained in \S\ref{back}. To compute the EM we have used coronal abundances of \cite{SchKS_2012} and standard ionization equilibrium given by CHIANTI \citep{DerML_1996, DelDY_2015}. The obtained EM-loci curves for all four rasters in the X-region are shown in Fig.~\ref{em_loci}. We have also plotted the histograms of number of crossings of EM curves within a temperature bin of $\log\,T[K] = $ 0.1 in the respective panels. In all the four plots, the left y-axis denote the EM values, whereas the right y-axis represents number of crossings in each temperature bin. 

The EM-loci curves along with the histograms suggest that the plasma within the X-shaped structure (enclosing the point P) during E2 is nearly isothermal with temperature within $\log\,T[K]=$ 6.10{--}6.20. In contrast, for E3 we note that the plasma is more multi-thermal in nature, with the histogram peaking between $\log\,T[K]=$ 6.10{--}6.30. The average temperatures in the X-region for all four rasters are noted in in Table~\ref{parameters}. 

For the P2-region, the EM-loci curves are shown in Fig.~\ref{em_p2}. It reveals that the peak formation temperature ranges between $\log\,T[K] = $ 6.20{--}6.30 (refer to Table~\ref{parameters}), which is somewhat larger than that noted in the X-region.

\paragraph{Plasma density and temperature diagnostics in the adjacent loops \newline} \label{loops}

\begin{figure*}[!htbp]
\centering
\includegraphics[width=1.0\textwidth]{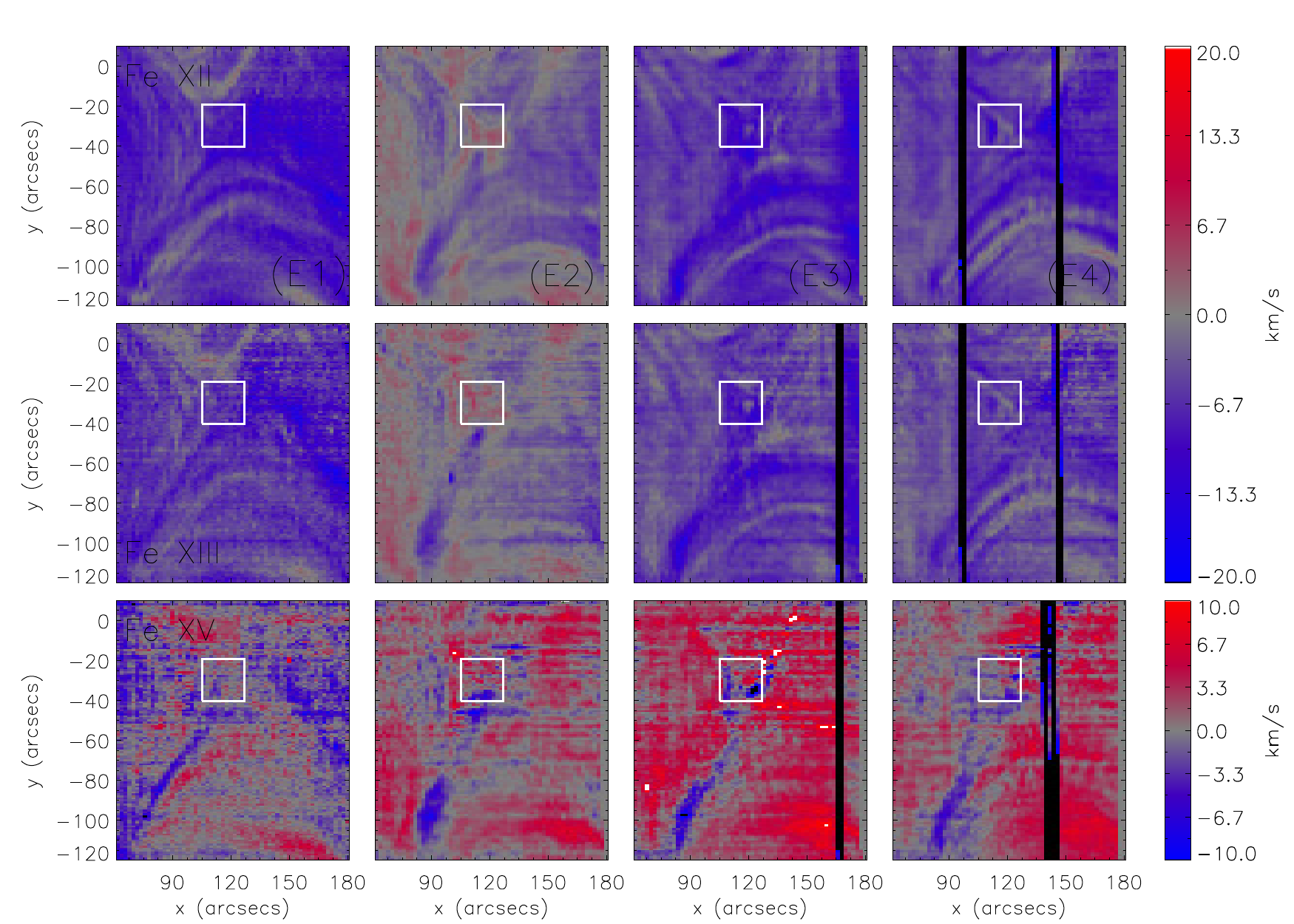}
\caption{Doppler maps corresponding to the X-region in \ion{Fe}{xii}, \ion{Fe}{xiii} and \ion{Fe}{xv} lines. E2 is closer to the disk center.}\label{dv2}
\end{figure*}

It is imperative to have a comparison of the electron densities and temperatures in the X-region (the point P is contained in this region) with those at the adjacent loops. Therefore, in Fig.~\ref{tels_loc} we have identified two additional regions- one on the TELs (shown in blue and identified as `L') and other at the loops belonging to the active region in the southern hemisphere (shown in black, identified as `D') of the X-region. The background is \ion{Fe}{xii} 195~{\AA} intensity map corresponding to E2. The average densities obtained at L, in most cases, is higher than those obtained at P. The former, in turn, is always less than those in D (see Table~\ref{parameters}). We note that densities in the TELs obtained here using \ion{Fe}{xii} lines is about an order of magnitude lower than those obtained by \cite{LiuWL_2011} using the line ratios of \ion{Si}{x} observed by CDS. \newline
In Fig.~\ref{tels_em}, we also plot the EM Loci curves obtained for locations L and D for E2 raster period. The plots show that the plasma at location D is more multi-thermal than location L. We further note that the temperature for TELs and loops corresponding to the active region in the southern hemisphere are somewhat higher (and more multi-thermal) than those obtained P i.e., within the X-shaped region (see Table~\ref{parameters}). The temperatures obtained for locations L and D are in agreement with those obtained for TELs by \citet{SheBB_1975, DelA_1999, GloHM_2003, Pev_2004, BalPN_2005}. This temperature range is maintained throughout the four rasters. 

\subsubsection{Doppler velocities and Spectral Line Width} \label{dv}

\begin{figure*}[!htbp]
\centering
\includegraphics[width=1.0\textwidth]{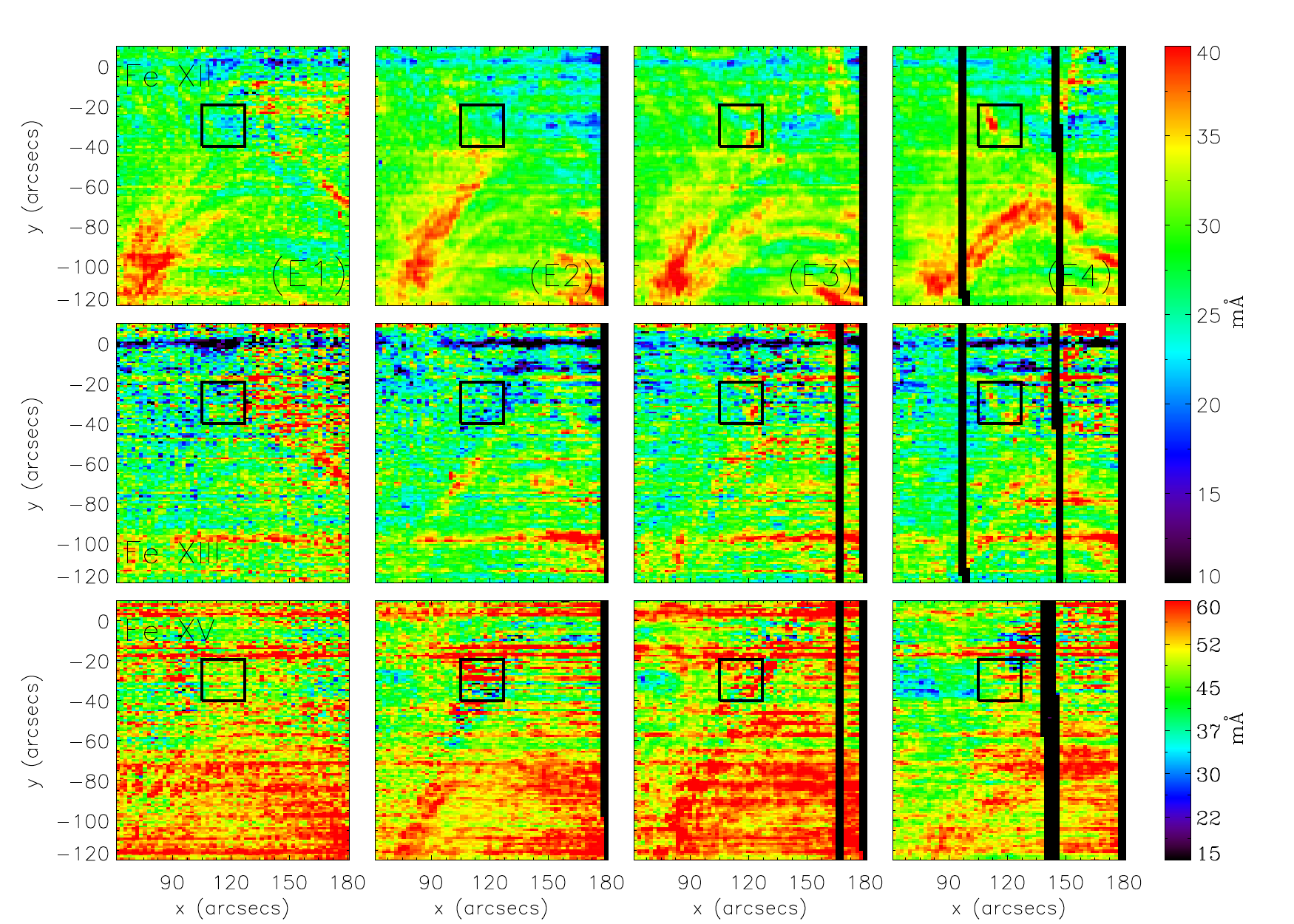}
\caption{FWHM maps in \ion{Fe}{xii}, \ion{Fe}{xiii} and \ion{Fe}{xv} lines corresponding to the X-region.}\label{nth}
\end{figure*}

We obtain the Doppler velocity and line width maps corresponding to the X-region for all four rasters in \ion{Fe}{xii}, \ion{Fe}{xiii} and \ion{Fe}{xv}. The maps are displayed in Figs.~\ref{dv2} \& \ref{nth}. The four columns represent the four EIS rasters. The over-plotted white (black) boxes in Doppler (line width) maps correspond to the blue boxes in Figs.~\ref{eis_int_1} \& \ref{eis_int_2}, enclosing the X-shaped structure. 

The Doppler velocity maps for \ion{Fe}{xii} and \ion{Fe}{xiii} are very similar. During the second raster E2, when the region is closest to the disk center, the X-shaped region shows dominant redshifts of $\sim$5{--}8~km~s$^{-1}$. On the contrary, during rasters E1 and E3, the same region shows predominant blueshifts of $\sim$10~km~s$^{-1}$. For the raster E4 when the active regions is farthest from the disk center, the LOS velocities are predominantly upflows ($\sim$5{--}8~km~s$^{-1}$). At higher temperature mapped by \ion{Fe}{xv} line, a significant fraction of the pixels is observed to have blueshifts (E1), redshifts (E2, E3) and very close to zero (E4) velocities. The average Doppler velocities in the box region are noted in Table~\ref{parameters}. Taking into account an uncertainty of $\sim$5~km~s$^{-1}$ in EIS velocity measurements \citep[see e.g.,][]{YouOM_2012}, we note that the direction of plasma flows along the LOS do not change.

The line width maps shown in Fig.~\ref{nth} are essentially the Full-Width-Half-Maxima (FWHM) obtained after subtracting the instrumental width \citep[56~m{\AA};][]{DosWM_2008}. In all the maps, the loops at the bottom have relatively higher FWHMs, but are best identified in \ion{Fe}{xii}. The FWHMs obtained within the box is $\sim$0.03~{\AA} for \ion{Fe}{xii} and \ion{Fe}{xiii} lines, for all four rasters (see, Table~\ref{parameters}), which is equivalent to the non-thermal velocity of $\sim$27~km~s$^{-1}$. To derive the non-thermal velocities, we have used the temperature in the X-region to be around 1.6~MK ($\log\,T[K]=$ 6.20) as was obtained using the EM-Loci procedure in \ref{temp}. However, there are a few individual pixels with FWHM as large as 0.068~{\AA} with an equivalent non-thermal velocity of $\sim$ 62~km~s$^{-1}$.  

The \ion{Fe}{xv} line has an average FWHM of $\sim$0.048~{\AA} ($\sim$30~km~s$^{-1}$) in X-shaped structure for all four rasters. There are pixels within the box, where the FWHM is as large as 0.18~{\AA}, which is equivalent to 115~km~s$^{-1}$ (assuming $\log\,T[K]=$ 6.2).

The average Doppler velocity and non-thermal width in P2 region are also noted in Table~\ref{parameters} and seen to be fairly similar to those in the X-region (within the error limits). In addition, the other plasma parameters in P2 (averages) are similar to those in the X-region. Therefore, we conclude that the X-region well represents the reconnection region, for all events.

\section{Summary and Conclusions} \label{summary}

This piece of study combines observations recorded by the AIA and the HMI on-board SDO and EIS on-board Hinode to study the formation, dynamics and plasma parameters of the trans-equatorial loops (TELs). The TELs were observed between pre-existing \textsl{AR~12230} in the southern hemisphere and the newly emerging \textsl{AR~12234} in the northern hemisphere. We have performed a comprehensive study of the evolution of TELs using AIA observations. The physical plasma parameters such as density, temperature, Doppler and non-thermal velocities in the loops as well as the interaction region between those loops are studied using EIS spectroscopic observations.

The observations recorded by AIA reveal that initially, loops belonging to the individual ARs evolve. These loops come closer to each other and form the X-shaped topology, leading to the formation of the TELs. This is highlighted in Fig.~\ref{reform} and is suggestive of the occurrence of the process of magnetic reconnection. The xt-plots in Fig.~\ref{xt} show that the inflow speed for the reconnection is 1{--}2~km~s$^{-1}$ whereas the outflow speed is $\sim$2~km~s$^{-1}$. These values are in close comparison to that of \cite{YokAM_2001, LiuWL_2011} in a flaring TEL system ($\sim$5~km~s$^{-1}$). However, we note that these are projected velocities measured in the plane-of-the-sky, hence gives the lower limit on the estimates. 

The intensity maps corresponding to the four EIS raster periods are derived using various spectral lines formed between $\log\, T[K] =$ 5.80{--}6.35. At lower temperatures, corresponding to \ion{Fe}{viii} to \ion{Fe}{x}, the X-region is seen very clearly with the adjacent loops being very bright. The intensity images of \ion{Fe}{xii} (formed at $\log\,T[K] \leq$ 6.20), show bright X-regions whereas at \ion{Fe}{xv} ($\log\,T[K]=$ 6.35), the X-region appears to be filled with plasma at E1 but completely void at E2 and partially filled at E3 and E4. Such darkening at the reconnection site has been attributed to density diminution \citep{DelA_1999, TriSMW_2007, SunCD_2015}. This result is reinforced by plotting the light curves in Fig.~\ref{lc}. \cite{YokAM_2001} reported similar voids in soft X-ray observations of magnetic reconnection events leading to flares. 

Within the X-region for all the four EIS raster periods, the electron densities are maintained steadily at $\log\, N_{e}=$ 8.46 to 8.67 cm$^{-3}$ for the \ion{Fe}{xii} and \ion{Fe}{xiii} lines. The EM-Loci curves suggest that the plasma is very nearly isothermal at $\log\, T[K]=$ 6.20 (i.e. 1.6~MK) within the X-region, which is somewhat larger than those reported by \cite{LiuWL_2011} at the cusp region (1.3~MK). However, \cite{SunCD_2015} found higher temperatures, between 1{--}5~MK at the magnetic reconnection site between TELs, but with an associated flare. Also, there are hints of multithermality of plasma in the X-region at E3. 

High cadence observations, at an interval of 12 seconds, further, reveal intermittent brightenings occurring close to the X-region (marked as P2 in Fig.~\ref{dots}). Coincidentally, such brightenings occurring during E2 phase of EIS raster observation were captured exactly by four exposures. This gave us the opportunity to understand the plasma parameters in this brightening point. Density estimates at P2 show that it has an intermediate value for \ion{Fe}{xii} line. Further, the plasma at P2 has slightly higher temperature, $\log\, T[K]=$ 6.20{--}6.30. We emphasize that we have not incorporated background/foreground emission for temperature estimates for reasons explained in \S\ref{back}.

We have also compared the densities and temperatures obtained in the X-region with those obtained in TELs and other AR loops. The average electron densities in the TELs (location L of Fig.~\ref{tels_loc}) exceed those in the X-region for all the four rasters. Similarly, the electron densities at location D is larger than those in the TELs. However, the densities at L is noted to be lower by an order of magnitude as reported by \cite[see, e.g.,][]{LiuWL_2011}, using the \ion{Si}{x} line of CDS. We also find that just beneath the X-region, a set of loops persist at $\sim$2~MK and the TELs are at intermediate temperatures. Existence of such hot loops neighbouring the X-region has also been reported by \cite{DelA_1999, GloHM_2003, Pev_2004, BalPN_2005}, though in the presence of eruptive events like flares or coronal mass ejections. The peak formation temperature in the loops belonging to the ARs individually is similar to that obtained at P2.  

The Doppler velocity maps show a mixture of upflows and downflows for the comparatively cooler spectral lines in the X-region. Near the disk center position, \ion{Fe}{xii}, \ion{Fe}{xiii} and \ion{Fe}{xv} show strong downflows but a mixture of zero velocities and blueshifts away from the center. These flows are $\sim$5{--}8~km~s$^{-1}$ upflows/downflows depending on the raster period. We note that the loops emanating from the reconnection show similar magnitude of Doppler velocities. With off-limb observations, \cite{HarMD_2003, Bro_2006, LiuWL_2011} have also reported similar bidirectional flows in such loops. The on-disk LOS velocities form the third mutually orthogonal component to the outflow speeds at the X-region noted above.

In the X-region, the average FWHM is 0.03~{\AA} which translates to about 27~km~s$^{-1}$ with the maximum being $\sim$ 62~km~s$^{-1}$ (corresponding to the brightest pixels) for \ion{Fe}{xii} and \ion{Fe}{xiii}. However, for \ion{Fe}{xv}, the X-region show a FWHM of $\sim$0.05~{\AA} translating to $\sim$30~km~s$^{-1}$ whereas the maximum is 0.18~{\AA}, equivalent to $\sim$115~km~s$^{-1}$. It is interesting to note that in case of \ion{Fe}{xv}, the enhanced FWHM region coincides with the dark X-region in intensity map. We note that the non-thermal velocities are obtained considering the temperature of the X-region ($\log\,T[K]=$ 6.20). The Doppler velocities and FWHM in P2 region is also comparable to those in the X-region. We highlight that the temperatures used for these estimates are representative of electron temperature of the plasma which can be considerably different than ion temperatures. This is particularly true in regions undergoing magnetic reconnection where equilibrium conditions are no longer valid. Considering that the ion temperatures are generally larger than the electron temperatures, it is apparent that the non-thermal velocities are somewhat overestimated here \citep{SheFS_1997, TuMW_1998, Lan_2007}. 

The results obtained based on the xt-plots derived using the AIA observations combined with those from the EIS suggests that the TELs formed through the process of reconnection at the X-region formed between the two active regions. This study, for the first time, provides measurements of plasma parameters such as electron density, temperature, Doppler shifts as well as non-thermal velocities at X-region. We interpret that this is an example of homologous low-intensity magnetic reconnections occurring in TELs where the energy released is predominantly used in plasma flows and as kinetic energy source for the field lines snapping away and getting reoriented in some other direction. The intensity increment observed in AIA 193~{\AA} images is very small but not accompanied by any flares or coronal mass ejections, unlike those reported by \cite{KhaH_2000, BalPN_2005}. It is plausible that such small-scale reconnections of TELs steadily keep feeding energy into the solar corona, thereby being a source of heating. However, further studies are required to confirm how frequent such loop systems are in the solar atmosphere in a given time interval and estimation of energy released by them. The physical plasma parameters obtained in this study may provide constraints for MHD simulations of magnetic reconnections in future.

\begin{acknowledgements} 
We thank the Referee for reading the manuscript carefully and provide valuable comments. We also thank Peter R. Young  for various discussions and S. K. Solanki for his comments on an earlier version of the manuscript. This research is supported by the Max-Planck India Partner Group of MPS at IUCAA that is funded by MPG and DST. AIA and HMI data are courtesy of SDO (NASA). Facilities: SDO (AIA). Hinode is a Japanese mission developed and launched by ISAS/JAXA, collaborating with NAOJ as a domestic partner, NASA and STFC (UK) as international partners. Scientific operation of the Hinode mission is conducted by the Hinode science team organized at ISAS/JAXA. This team mainly consists of scientists from institutes in the partner countries. Support for the post-launch operation is provided by JAXA and NAOJ (Japan), STFC (UK), NASA, ESA, and NSC (Norway).
\end{acknowledgements}

\bibliographystyle{aa}
\bibliography{references}

\clearpage
\thispagestyle{empty}

\begin{sidewaystable*}
\vspace{1cm}
\caption{Intensities averaged over the X-region (`P'), TELs (`L'), hot loops beneath the X-region (`D'), the bright region hosting recurrent brightenings (`P2') and region considered for background/ foreground assessment (`B'), respectively in Fig.~\ref{tels_loc}. The averaged intensities for \ion{Mg}{7} and \ion{Si}{7} are not noted here because the background averaged intensities exceed those in the region(s) under consideration for all four EIS observations.}\label{inten}
\vspace{1cm}
\begin{adjustwidth}{-2.0cm}{}
\normalsize
\begin{tabular}{| c | c | c | c | c | c | c | c | c | c | c | c | c | c | c | c | c | c | c | c | c |} \hline
\multicolumn{1}{|c|}{Fe lines} &\multicolumn{5}{c|}{E1} &\multicolumn{5}{c|}{E2} &\multicolumn{5}{c|}{E3} &\multicolumn{5}{c|}{E4} \\ \hline

\multicolumn{1}{|c|}{} &\multicolumn{1}{c|}{P} &\multicolumn{1}{c|}{L} &\multicolumn{1}{c|}{D} &\multicolumn{1}{c|}{P2} &\multicolumn{1}{c|}{B} &\multicolumn{1}{c|}{P} &\multicolumn{1}{c|}{L} &\multicolumn{1}{c|}{D} &\multicolumn{1}{c|}{P2} &\multicolumn{1}{c|}{B} &\multicolumn{1}{c|}{P} &\multicolumn{1}{c|}{L} &\multicolumn{1}{c|}{D} &\multicolumn{1}{c|}{P2} &\multicolumn{1}{c|}{B} &\multicolumn{1}{c|}{P} &\multicolumn{1}{c|}{L} &\multicolumn{1}{c|}{D} &\multicolumn{1}{c|}{P2} &\multicolumn{1}{c|}{B}\\ \hline

\multicolumn{1}{|c|}{\ion{Fe}{viii}} &\multicolumn{1}{|c|}{568.0} &\multicolumn{1}{c|}{420.4} &\multicolumn{1}{c|}{411.3} &\multicolumn{1}{c|}{-} &\multicolumn{1}{|c|}{568.0}  

 &\multicolumn{1}{|c|}{274.0} &\multicolumn{1}{c|}{597.3} &\multicolumn{1}{c|}{498.4} &\multicolumn{1}{c|}{136.6} &\multicolumn{1}{c|}{128.4}

 &\multicolumn{1}{|c|}{165.8} &\multicolumn{1}{c|}{482.2} &\multicolumn{1}{c|}{411.8} &\multicolumn{1}{c|}{-} &\multicolumn{1}{c|}{1335.3}

 &\multicolumn{1}{|c|}{488.0}  &\multicolumn{1}{c|}{421.8} &\multicolumn{1}{c|}{799.9} &\multicolumn{1}{c|}{-} &\multicolumn{1}{c|}{5273.9}\\ \hline

\multicolumn{1}{|c|}{\ion{Fe}{ix}} &\multicolumn{1}{|c|}{69.2} &\multicolumn{1}{c|}{190.2} &\multicolumn{1}{c|}{124.3} &\multicolumn{1}{c|}{-} &\multicolumn{1}{c|}{46.3} 

&\multicolumn{1}{|c|}{174.7}  &\multicolumn{1}{c|}{205.4} &\multicolumn{1}{c|}{129.5} &\multicolumn{1}{c|}{21.1} &\multicolumn{1}{c|}{362.6}

   &\multicolumn{1}{|c|}{117.6} &\multicolumn{1}{c|}{156.0} &\multicolumn{1}{c|}{130.8} &\multicolumn{1}{c|}{-} &\multicolumn{1}{c|}{85.3} 

 &\multicolumn{1}{|c|}{201.2}&\multicolumn{1}{c|}{185.8} &\multicolumn{1}{c|}{285.1} &\multicolumn{1}{c|}{-} &\multicolumn{1}{c|}{17.0}\\ \hline

\multicolumn{1}{|c|}{\ion{Fe}{x}} &\multicolumn{1}{|c|}{516.9} &\multicolumn{1}{c|}{1203.6} &\multicolumn{1}{c|}{1003.2} &\multicolumn{1}{c|}{-} &\multicolumn{1}{c|}{228.1} 

 &\multicolumn{1}{|c|}{1171.1} &\multicolumn{1}{c|}{957.3} &\multicolumn{1}{c|}{977.0} &\multicolumn{1}{c|}{512.4} &\multicolumn{1}{c|}{289.4} 

 &\multicolumn{1}{|c|}{802.7} &\multicolumn{1}{c|}{937.8} &\multicolumn{1}{c|}{964.2} &\multicolumn{1}{c|}{-} &\multicolumn{1}{c|}{1990.9} 

 &\multicolumn{1}{|c|}{962.6} &\multicolumn{1}{c|}{928.8} &\multicolumn{1}{c|}{1711.2} &\multicolumn{1}{c|}{-} &\multicolumn{1}{c|}{1312.1}\\ \hline

\multicolumn{1}{|c|}{\ion{Fe}{xi}}&\multicolumn{1}{|c|}{2096.8}  &\multicolumn{1}{c|}{4222.3} &\multicolumn{1}{c|}{4499.5} &\multicolumn{1}{c|}{-} &\multicolumn{1}{c|}{2114.4} 

 &\multicolumn{1}{|c|}{4382.8}  &\multicolumn{1}{c|}{3267.3} &\multicolumn{1}{c|}{4234.2} &\multicolumn{1}{c|}{1551.1} &\multicolumn{1}{c|}{6534.7}

&\multicolumn{1}{|c|}{3586.7} &\multicolumn{1}{c|}{3505.8} &\multicolumn{1}{c|}{4249.4} &\multicolumn{1}{c|}{-} &\multicolumn{1}{c|}{13723.4} 

&\multicolumn{1}{|c|}{3761.4} &\multicolumn{1}{c|}{3566.7} &\multicolumn{1}{c|}{5950.9} &\multicolumn{1}{c|}{-} &\multicolumn{1}{c|}{26928.2}\\ \hline

\multicolumn{1}{|c|}{\ion{Fe}{xii}} &\multicolumn{1}{|c|}{794.5} &\multicolumn{1}{c|}{1394.2} &\multicolumn{1}{c|}{1597.9} &\multicolumn{1}{c|}{-} &\multicolumn{1}{c|}{178.2} 

&\multicolumn{1}{|c|}{1132.8} &\multicolumn{1}{c|}{952.8} &\multicolumn{1}{c|}{1406.4} &\multicolumn{1}{c|}{610.7} &\multicolumn{1}{c|}{167.0} 

 &\multicolumn{1}{|c|}{1203.0}  &\multicolumn{1}{c|}{986.9} &\multicolumn{1}{c|}{1629.8} &\multicolumn{1}{c|}{-} &\multicolumn{1}{c|}{166.1} 

 &\multicolumn{1}{|c|}{1110.1} &\multicolumn{1}{c|}{1217.3} &\multicolumn{1}{c|}{1951.2} &\multicolumn{1}{c|}{-} &\multicolumn{1}{c|}{170.2}\\ 
 
\multicolumn{1}{|c|}{192~{\AA}} &\multicolumn{1}{|c|}{} &\multicolumn{1}{c|}{} &\multicolumn{1}{c|}{} &\multicolumn{1}{c|}{} &\multicolumn{1}{c|}{} 

&\multicolumn{1}{|c|}{} &\multicolumn{1}{c|}{} &\multicolumn{1}{c|}{} &\multicolumn{1}{c|}{} &\multicolumn{1}{c|}{} 

 &\multicolumn{1}{|c|}{}  &\multicolumn{1}{c|}{} &\multicolumn{1}{c|}{} &\multicolumn{1}{c|}{} &\multicolumn{1}{c|}{} 

 &\multicolumn{1}{|c|}{} &\multicolumn{1}{c|}{} &\multicolumn{1}{c|}{} &\multicolumn{1}{c|}{} &\multicolumn{1}{c|}{}\\  \hline

\multicolumn{1}{|c|}{\ion{Fe}{xii}} &\multicolumn{1}{|c|}{2804.1} &\multicolumn{1}{c|}{4832.1} &\multicolumn{1}{c|}{5484.3} &\multicolumn{1}{c|}{-} &\multicolumn{1}{c|}{638.8} 

 &\multicolumn{1}{|c|}{4095.9} &\multicolumn{1}{c|}{3504.4} &\multicolumn{1}{c|}{4976.8} &\multicolumn{1}{c|}{2141.3} &\multicolumn{1}{c|}{592.2} 

 &\multicolumn{1}{|c|}{4165.0} &\multicolumn{1}{c|}{3547.3} &\multicolumn{1}{c|}{5607.1} &\multicolumn{1}{c|}{-} &\multicolumn{1}{c|}{593.9} 

 &\multicolumn{1}{|c|}{3934.5} &\multicolumn{1}{c|}{4297.9} &\multicolumn{1}{c|}{6630.3} &\multicolumn{1}{c|}{-} &\multicolumn{1}{c|}{619.4}\\ 
 
 \multicolumn{1}{|c|}{195~{\AA}} &\multicolumn{1}{|c|}{} &\multicolumn{1}{c|}{} &\multicolumn{1}{c|}{} &\multicolumn{1}{c|}{} &\multicolumn{1}{c|}{} 

&\multicolumn{1}{|c|}{} &\multicolumn{1}{c|}{} &\multicolumn{1}{c|}{} &\multicolumn{1}{c|}{} &\multicolumn{1}{c|}{} 

 &\multicolumn{1}{|c|}{}  &\multicolumn{1}{c|}{} &\multicolumn{1}{c|}{} &\multicolumn{1}{c|}{} &\multicolumn{1}{c|}{} 

 &\multicolumn{1}{|c|}{} &\multicolumn{1}{c|}{} &\multicolumn{1}{c|}{} &\multicolumn{1}{c|}{} &\multicolumn{1}{c|}{}\\  \hline

\multicolumn{1}{|c|}{\ion{Fe}{xiii}} &\multicolumn{1}{|c|}{1688.0} &\multicolumn{1}{c|}{2412.1} &\multicolumn{1}{c|}{3123.3} &\multicolumn{1}{c|}{-} &\multicolumn{1}{c|}{269.8} 

&\multicolumn{1}{|c|}{1211.6}  &\multicolumn{1}{c|}{1702.9} &\multicolumn{1}{c|}{2535.9} &\multicolumn{1}{c|}{1098.5} &\multicolumn{1}{c|}{210.4} 

 &\multicolumn{1}{|c|}{1946.6}   &\multicolumn{1}{c|}{1774.3} &\multicolumn{1}{c|}{3345.8} &\multicolumn{1}{c|}{-} &\multicolumn{1}{c|}{215.0}

 &\multicolumn{1}{|c|}{1628.9} &\multicolumn{1}{c|}{2129.9} &\multicolumn{1}{c|}{3467.5} &\multicolumn{1}{c|}{-} &\multicolumn{1}{c|}{218.1}\\ \hline

\multicolumn{1}{|c|}{\ion{Fe}{xiv}} &\multicolumn{1}{|c|}{695.3}&\multicolumn{1}{c|}{738.8} &\multicolumn{1}{c|}{1018.3} &\multicolumn{1}{c|}{-} &\multicolumn{1}{c|}{159.8} 

 &\multicolumn{1}{|c|}{402.8} &\multicolumn{1}{c|}{660.8} &\multicolumn{1}{c|}{942.3} &\multicolumn{1}{c|}{512.1} &\multicolumn{1}{c|}{69.7} 

 &\multicolumn{1}{|c|}{573.5}  &\multicolumn{1}{c|}{684.6} &\multicolumn{1}{c|}{1114.8} &\multicolumn{1}{c|}{-} &\multicolumn{1}{c|}{193.3} 

 &\multicolumn{1}{|c|}{538.0} &\multicolumn{1}{c|}{737.3} &\multicolumn{1}{c|}{981.0} &\multicolumn{1}{c|}{-} &\multicolumn{1}{c|}{53.9}\\ \hline

\multicolumn{1}{|c|}{\ion{Fe}{xv}} &\multicolumn{1}{|c|}{4840.5} &\multicolumn{1}{c|}{3927.6} &\multicolumn{1}{c|}{6162.3} &\multicolumn{1}{c|}{-} &\multicolumn{1}{c|}{1183.1} 

 &\multicolumn{1}{|c|}{1403.2}  &\multicolumn{1}{c|}{4113.0} &\multicolumn{1}{c|}{6550.6} &\multicolumn{1}{c|}{3175.2} &\multicolumn{1}{c|}{451.3}

 &\multicolumn{1}{|c|}{2280.3} &\multicolumn{1}{c|}{4854.3} &\multicolumn{1}{c|}{6757.8} &\multicolumn{1}{c|}{-} &\multicolumn{1}{c|}{480.2}

 &\multicolumn{1}{|c|}{2667.8} &\multicolumn{1}{c|}{4345.2} &\multicolumn{1}{c|}{4909.9}  &\multicolumn{1}{c|}{-} &\multicolumn{1}{c|}{448.8}\\ \hline

\end{tabular} 
\end{adjustwidth}
\end{sidewaystable*}
\clearpage

\clearpage
\thispagestyle{empty}
\begin{sidewaystable*}
\begin{center}
\vspace{1cm}
\caption{Plasma parameters for different locations in the FOV using different EIS spectral lines. These are computed over the white (enclosing the point `P' of panel A in Fig.~\ref{xt}), blue (TELs,  marked `L') and black (loops at the bottom,  marked `D') boxes marked in Fig.~\ref{tels_loc}. Similar computations are done for `P2' region, which undergoes concurrent intensity enhancements during E2 (as shown in Fig.~\ref{dots}). The density estimates consider background/ foreground emission over the region marked `B' in the Fig.~\ref{tels_loc}. No background/ foreground emission treatment has been done for the temperature estimates.}\label{parameters}
\vspace{1cm}
\normalsize
\begin{tabular}{| c | c | c | c | c | c | c | c | c | c | c | c | c |} \hline
\multicolumn{1}{|c|}{Parameter} &\multicolumn{3}{c|}{E1} &\multicolumn{3}{c|}{E2} &\multicolumn{3}{c|}{E3} &\multicolumn{3}{c|}{E4} \\ \hline

\multicolumn{1}{|c|}{$\log\, N_{e}$ (cm$^{-3}$)} &\multicolumn{3}{c|}{} &\multicolumn{3}{c|}{} &\multicolumn{3}{c|}{} &\multicolumn{3}{c|}{}  \\

\multicolumn{1}{|c|}{\ion{Fe}{xii} } &\multicolumn{3}{c|}{} &\multicolumn{3}{c|}{} &\multicolumn{3}{c|}{} &\multicolumn{3}{c|}{} \\ 

\multicolumn{1}{|c|}{P} &\multicolumn{3}{c|}{8.67} &\multicolumn{3}{c|}{8.56} &\multicolumn{3}{c|}{8.46} &\multicolumn{3}{c|}{8.54}\\ \hline

\multicolumn{1}{|c|}{L} &\multicolumn{3}{c|}{8.72} &\multicolumn{3}{c|}{8.61} &\multicolumn{3}{c|}{8.46} &\multicolumn{3}{c|}{8.55}\\ \hline

\multicolumn{1}{|c|}{D} &\multicolumn{3}{c|}{8.80} &\multicolumn{3}{c|}{8.77} &\multicolumn{3}{c|}{8.67} &\multicolumn{3}{c|}{8.79}\\  \hline

\multicolumn{1}{|c|}{P2} &\multicolumn{3}{c|}{-} &\multicolumn{3}{c|}{8.50} &\multicolumn{3}{c|}{-} &\multicolumn{3}{c|}{-}\\ \hline

\multicolumn{1}{|c|}{$\log\, T[/K]$} &\multicolumn{3}{c|}{} &\multicolumn{3}{c|}{} &\multicolumn{3}{c|}{} &\multicolumn{3}{c|}{} \\ 

\multicolumn{1}{|c|}{P} &\multicolumn{3}{c|}{6.25} &\multicolumn{3}{c|}{6.15} &\multicolumn{3}{c|}{6.25} &\multicolumn{3}{c|}{6.25} \\ \hline

\multicolumn{1}{|c|}{L} &\multicolumn{3}{c|}{6.15} &\multicolumn{3}{c|}{6.15} &\multicolumn{3}{c|}{6.15} &\multicolumn{3}{c|}{6.15} \\ \hline

\multicolumn{1}{|c|}{D} &\multicolumn{3}{c|}{6.25} &\multicolumn{3}{c|}{6.25} &\multicolumn{3}{c|}{6.25} &\multicolumn{3}{c|}{6.25} \\  \hline

\multicolumn{1}{|c|}{P2} &\multicolumn{3}{c|}{-} &\multicolumn{3}{c|}{6.25} &\multicolumn{3}{c|}{-} &\multicolumn{3}{c|}{-} \\ \hline

\multicolumn{1}{|c|}{Ion name} &\multicolumn{1}{|c|}{\ion{Fe}{xii}} &\multicolumn{1}{|c|}{\ion{Fe}{xiii}} &\multicolumn{1}{|c|}{\ion{Fe}{xv}}  &\multicolumn{1}{|c|}{\ion{Fe}{xii}} &\multicolumn{1}{|c|}{\ion{Fe}{xiii}} &\multicolumn{1}{|c|}{\ion{Fe}{xv}} &\multicolumn{1}{|c|}{\ion{Fe}{xii}} &\multicolumn{1}{|c|}{\ion{Fe}{xiii}} &\multicolumn{1}{|c|}{\ion{Fe}{xv}} &\multicolumn{1}{|c|}{\ion{Fe}{xii}} &\multicolumn{1}{|c|}{\ion{Fe}{xiii}} &\multicolumn{1}{|c|}{\ion{Fe}{xv}}\\ 
\hline

\multicolumn{1}{|c|}{Doppler velocity (~km~s$^{-1}$) at P} &\multicolumn{1}{|c|}{-8.08} &\multicolumn{1}{|c|}{-6.15} &\multicolumn{1}{|c|}{-0.37}  &\multicolumn{1}{|c|}{0.10} &\multicolumn{1}{|c|}{1.23} &\multicolumn{1}{|c|}{0.29} &\multicolumn{1}{|c|}{-8.53} &\multicolumn{1}{|c|}{-7.31} &\multicolumn{1}{|c|}{-0.25} &\multicolumn{1}{|c|}{-5.94} &\multicolumn{1}{|c|}{-4.82} &\multicolumn{1}{|c|}{-0.37}\\ \hline

\multicolumn{1}{|c|}{Doppler velocity (~km~s$^{-1}$) at P2} &\multicolumn{1}{|c|}{-} &\multicolumn{1}{|c|}{-} &\multicolumn{1}{|c|}{-}  &\multicolumn{1}{|c|}{-2.22} &\multicolumn{1}{|c|}{-1.69} &\multicolumn{1}{|c|}{2.82} &\multicolumn{1}{|c|}{-} &\multicolumn{1}{|c|}{-} &\multicolumn{1}{|c|}{-} &\multicolumn{1}{|c|}{-} &\multicolumn{1}{|c|}{-} &\multicolumn{1}{|c|}{-}\\ \hline

\multicolumn{1}{|c|}{FWHM (m{\AA}) at P} &\multicolumn{1}{|c|}{0.026} &\multicolumn{1}{|c|}{0.027} &\multicolumn{1}{|c|}{0.048}  &\multicolumn{1}{|c|}{0.028} &\multicolumn{1}{|c|}{0.025} &\multicolumn{1}{|c|}{0.048} &\multicolumn{1}{|c|}{0.029} &\multicolumn{1}{|c|}{0.028} &\multicolumn{1}{|c|}{0.050} &\multicolumn{1}{|c|}{0.031} &\multicolumn{1}{|c|}{0.028} &\multicolumn{1}{|c|}{0.046}\\ 
\multicolumn{1}{|c|}{(excluding instrumental width)} &\multicolumn{1}{|c|}{} &\multicolumn{1}{|c|}{} &\multicolumn{1}{|c|}{}  &\multicolumn{1}{|c|}{} &\multicolumn{1}{|c|}{} &\multicolumn{1}{|c|}{} &\multicolumn{1}{|c|}{} &\multicolumn{1}{|c|}{} &\multicolumn{1}{|c|}{} &\multicolumn{1}{|c|}{} &\multicolumn{1}{|c|}{} &\multicolumn{1}{|c|}{}\\ \hline

\multicolumn{1}{|c|}{FWHM (m{\AA}) at P2} &\multicolumn{1}{|c|}{-} &\multicolumn{1}{|c|}{-} &\multicolumn{1}{|c|}{-}  &\multicolumn{1}{|c|}{0.033} &\multicolumn{1}{|c|}{0.030} &\multicolumn{1}{|c|}{0.039} &\multicolumn{1}{|c|}{-} &\multicolumn{1}{|c|}{-} &\multicolumn{1}{|c|}{-} &\multicolumn{1}{|c|}{-} &\multicolumn{1}{|c|}{-} &\multicolumn{1}{|c|}{-}\\ 
\multicolumn{1}{|c|}{(excluding instrumental width)} &\multicolumn{1}{|c|}{} &\multicolumn{1}{|c|}{} &\multicolumn{1}{|c|}{}  &\multicolumn{1}{|c|}{} &\multicolumn{1}{|c|}{} &\multicolumn{1}{|c|}{} &\multicolumn{1}{|c|}{} &\multicolumn{1}{|c|}{} &\multicolumn{1}{|c|}{} &\multicolumn{1}{|c|}{} &\multicolumn{1}{|c|}{} &\multicolumn{1}{|c|}{}\\
\hline

\end{tabular} 
\end{center}
\end{sidewaystable*}
\clearpage



\end{document}